\documentclass[referee]{jfm}
\usepackage{epsfig}
\usepackage{times}
\usepackage{amsmath,amsfonts,amssymb,amstext,amscd,array}
\usepackage{graphicx}
\usepackage{bm}
\usepackage{natbib}

   \newcommand{\rem}[1]{}
\newcommand{\grad}{{\bm \nabla}}
\newcommand{\gradH}{{\bm \nabla}_{\! H}}
\newcommand{\bfk}{\mathbf{k}}
\newcommand{\bfr}{\mathbf{r}}
\newcommand{\bfu}{\mathbf{u}}
\newcommand{\bfv}{\mathbf{v}}
\newcommand{\bfx}{\mathbf{x}}
\newcommand{\bfO}{\mathbf{0}}
\newcommand{\eps}{\epsilon}
\newcommand{\rhot}{\tilde{\rho}}
\newcommand{\za}{\alpha}
\newcommand{\w}{\omega}
\newcommand{\sla}{\mbox{\sl a}}
\newcommand{\slv}{\mbox{\sl v}}
\newcommand{\dd}[2]{\frac{\partial #1}{\partial #2}}
\newcommand{\avez}[1]{\langle #1 \rangle_z}

\newcommand{\RR}{{\mathbb{R}}}

\begin{document}
\title[Low Rossby limiting dynamics]{Low Rossby limiting dynamics for
  stably stratified flow with Finite Froude number}
\author[Beth A. Wingate, Pedro Embid,  Miranda Holmes-Cerfon and Mark A. Taylor]
{B\ls E\ls T\ls H\ns A.\ns W\ls I\ls N\ls G\ls A\ls T\ls E\ls 
  \thanks{MS D413, Los Alamos National Laboratory,
    Los Alamos, NM 87544},
\ns P\ls E\ls D\ls R\ls O\ls \ns E\ls M\ls B\ls I\ls D\ls
  \thanks{Department of Mathematics and Statistics, the University of New Mexico,
          Albuquerque, NM 87131}\break
\ns M\ls I\ls R\ls A\ls N\ls D\ls A\ls \ns H\ls O\ls L\ls M\ls E\ls
S-C\ls E\ls R\ls F\ls O\ls N\ls  \ls
   \thanks{New York University, Courant Institute of Mathematical Sciences, New York, NY 10012 }
\ns A\ls N\ls D\ls
\ns M\ls A\ls R\ls K\ns A.\ns T\ls A\ls Y\ls L\ls O\ls R\ls
   \thanks{Sandia National Laboratories, Albuquerque, NM 87185} } 
\maketitle
\bibliographystyle{jfm}
\begin{abstract}
  
  In this paper we explore the strong rotation limit of the rotating
  and stratified Boussinesq equations with periodic boundary
  conditions when the stratification is order one ([Rossby number] $Ro
  = \epsilon$, [Froude number] $Fr = O(1)$, as $\epsilon \rightarrow
  0$).  Using the same framework of \cite{EmbidPF:LowFnl} we show that
  the slow dynamics decouples from the fast. Furthermore, we derive
  equations for the slow dynamics and their conservation laws. The
  horizontal momentum equations reduce to the two-dimensional
  Navier-Stokes equations. The equation for the vertically averaged
  vertical velocity includes a term due to the vertical average of the
  buoyancy. The buoyancy equation, the only variable to retain its
  three-dimensionality, is advected by all three two-dimensional slow
  velocity components. The conservation laws for the slow dynamics
  include those for the two-dimensional Navier Stokes equations and a
  new conserved quantity that describes dynamics between the vertical
  kinetic energy and the buoyancy. The leading order potential
  enstrophy is slow while the leading order total energy retains both
  fast and slow dynamics. We also perform forced numerical simulations
  of the rotating Boussinesq equations to demonstrate support for
  three aspects of the theory in the limit $Ro \rightarrow 0$: 1) we
  find the formation and persistence of large-scale columnar
  Taylor-Proudman flows in the presence of $O(1)$ Froude number; after
  a spin-up time 2) the ratio of the slow total energy to the total
  energy approaches a constant and that at the smallest Rossby numbers
  that constant approaches one; and 3) the ratio of the slow potential
  enstrophy to the total potential enstrophy also approaches a
  constant and that at the lowest Rossby numbers that constant is one.
  The results of the numerical simulations indicate that even in the
  presence of the low wave number white noise forcing the dynamics
  exhibit characteristics of the theory.

\end{abstract}

\section{Introduction}
\label{sec:intro}
For planetary scale rotating and stratified fluid dynamics
\cite{CharneyJG:Onsam} estimated the orders of magnitude of different
terms in the Euler equations by using typical values of large scale
atmospheric motion.  From these arguments, which included approximate
hydrostatic and geostrophic balance, he derived reduced sets of
equations called the quasi-geostrophic equations (QG) that are widely
used in idealized studies of oceanic and atmospheric dynamics. In
addition to finding equations that govern the large scales Charney
also points out that the QG equations 'filtered out' the
inconsequential fast waves from the large scale motions.


In the work of \cite{EmbidPF:Aveofg, EmbidPF:LowFnl, MajdaAJ:Aveofg}
they showed that the QG limit ([Rossby number] $Ro \rightarrow 0$,
[Froude number] $Fr \rightarrow 0$, and $Fr/Ro = f/N$ finite), can also be
derived from the rotating Boussinesq equations with periodic boundary
conditions, Eqs. (\ref{eq:bousnd2})-(\ref{eq:buoynd2}).  Their
asymptotic analysis relies on the separation of fast and slow time
scales and incorporates the fast waves that were filtered out in
Charney's analysis. The resulting limiting equations are obtained by
averaging over the fast time scale and accounts for three-waves
interactions of fast and slow waves. Moreover, a rigorous
justification of this approach was given by a direct application of
Schochet's method of cancellation of oscillations for hyperbolic
equations, \cite{schochet1994}.  In these papers, taking $Ro
\rightarrow 0$ corresponds to geostrophic balance and taking $Fr
\rightarrow 0$ corresponds to hydrostatic balance.  When both these
parameters go to zero the equations for the slow dynamics decouple
from the fast and are Charney's QG equations.

In addition to the quasi-geostrophic regime described above we
consider the dynamics of two other dynamical regimes: 1) the strong
stratification limit where the physics is dominated by strong
stratification but has only weak rotational effects and 2) the strong
rotation limit where the physics is dominated by fast rotation but is
only weakly stratified.

The first limit, the strong stratification limit, is thought to be
important in geophysical fluid dynamics, see the review by
\cite{RileyJJ:Flumps}, because it describes flows that occur at length
scales between the large, quasi-geostrophic scales and the small
scales where energy is dissipated. Fluid dynamical theory for this
physical regime has been explored by \cite{RileyJJ:Flumps,
  RileyJJ:Dyntsi, Babin1996, BabinA:Onara, Babin1998, Babin2002,
  EmbidPF:LowFnl}. Parametrically this regime is described by $Fr
\rightarrow 0, ~Ro = O(1), ~f/N \rightarrow 0$. Here it has also been
found that the slow dynamics decouples from the fast and that it leads
to new equations for the slow dynamics that are not the QG equations
derived by Charney.

One way the slow dynamics of the QG limit differs from the slow
dynamics of the strong stratification limit is in the role of the zero
frequency dispersive waves. To explore this we examine the
eigenfrequencies of the nondimensional linearized rotating Boussinesq equations,
Eqs.  (\ref{eq:bousnd2})-(\ref{eq:buoynd2}) in the absence of dissipative effects,
\begin{equation}
\label{eq:bousdisp}
\omega(\mathbf{k}) = \pm \frac{(Fr^2 m^2 +
  Ro ^2 ~|\mathbf{k_H}|^2)^{1/2}}{Ro Fr |\mathbf{k}|}, 
\qquad
\omega(\mathbf{k}) = 0  ~~~\text{(double)},
\end{equation}
where $|\mathbf{k_H}| = k^2 + l^2$ with $k$ and $l$ the horizontal
wave numbers, $m$ the vertical wave number, and $|\mathbf{k}| = k^2 +
l^2 + m^2$. There are two kinds of eigenfrequencies. The first kind
are the slow vortical modes that have zero frequency for all
$\mathbf{k}$, and contribute to the potential vorticity. The second
kind are the dispersive waves that have non-zero frequency but make no
contribution to potential vorticity.  The latter are the familiar
inertia-gravity waves that are filtered from Charney's QG equations.
In the strong stratification limit (see
\cite{EmbidPF:LowFnl,BabinA:Onara}) where $Fr \rightarrow 0, Ro = O(1),
f/N \rightarrow 0$, the fast eigenfrequencies are,
\begin{equation}
\omega_{Fr}(\mathbf{k}) = \pm \frac{|\mathbf{k_H}|}{|\mathbf{k}|}, 
\qquad 
\omega(\mathbf{k}) = 0 ~~~\text{(double)}.
\end{equation}
Again there are two kinds of eigenfrequencies, the slow vortical modes
and the fast gravity waves. However, here the fast waves contribute to
the slow dynamics when $\mathbf{k_H} = 0$. That is, one of the wave
modes, corresponding to horizontal averages with zero potential
vorticity, has a slow component that resonates with the PV bearing
vortical modes. This manifests itself in the vertically sheared
horizontal dynamics (VSHF) mode introduced by \cite{EmbidPF:LowFnl}
and investigated by \cite{SmithLM:Gensls,MajdaGrote1997}.

In this work, we look at the strong rotation (low Rossby) limit where
geostrophic balance dominates but the flow is only weakly stratified.
These dynamics are parametrically described by the limit $Ro
\rightarrow 0, Fr = O(1), ~f/N \rightarrow \infty$. 

This physical regime is relevant in regions of the deep ocean where
stratification is weak but rotational effects are dominant. For
example, \cite{VanHaren2005} observe values of $N = 0 \pm .4f ~(2.5 <
f/N < \infty)$ in the deep Mediterranean Sea and argue that the
dynamics in those regions are driven by both weak stratification and
the horizontal components rotation. In fact, they observe nonhydrostatic motions with
vertical velocities of the same order of magnitude as the horizontal.
Another region of the world where strong rotation and weak
stratification have been observed is in the deep Arctic Ocean.
Measurements in the Beaufort Gyre by \cite{Timmermans2007,
  Timmermans2010} show $f/N \approx 2$ above 2600
meters and $f/N \approx \infty$ between the depths of
2600 and 3600m. One of the reasons these investigators give for
studying the deep Arctic is that in their 2002 pilot study they
discovered the dynamics to be unexpectedly active in the deep ocean.
Weak stratification in the deep ocean at high latitudes has been noted
for the North Atlantic and North Pacific in \cite{Emeryetal1984} where
they compute mean profiles of density and Brunt-V\"ais\"ala frequency;
in the deep waters of the Arctic Ocean by \cite{Jones1995}; and in the
Southern Ocean by \cite{Heywood2002}.  Furthermore, warm core eddies
with depths of 1000 meters or more have been observed in the Arctic by
\cite{Woodgate2001}.

In the limit of strong rotation and weak stratification ($Ro \rightarrow 0,
Fr = O(1), ~f/N \rightarrow \infty$)
the fast eigenfrequencies  are,
\begin{equation}
\omega_{Ro}(\mathbf{k}) = \pm \frac{|m|}{|\mathbf{k}|}, 
\qquad 
\omega(\mathbf{k}) = 0 ~~~\text{(double)}.
\end{equation}
Again, there are two kinds of frequencies corresponding to fast
inertial waves and slow PV modes. Also, in this limit the non PV
bearing modes make a contribution to the slow dynamics, but this time
it occurs when $m=0$, which corresponds to vertically averaged
dynamics, which we refer to as Taylor-Proudman dynamics.

By using the general framework developed in \cite{EmbidPF:LowFnl} we
show that in the low Rossby number limit the horizontal and vertical
dynamics decouple. In the horizontal the slow equations are the
two-dimensional Navier-Stokes equations along with two conservation
laws, the horizontal kinetic energy and the vertical vorticity. In the
case where the flow is not stratified this is consistent with other
work \cite{QiaoningChen:Resirh}. The vertically averaged vertical
momentum equation is an advection-diffusion like equation that couples
to the buoyancy through its vertical average.  The slow equation for
the buoyancy is the only quantity that remains fully three dimensional
and is advected by all 3 components of the slow velocity.  The slow
equations for the vertical momentum and the buoyancy are coupled and
give rise to new conservation laws for the coupled $w-\rho$ dynamics.
We also show that the slow modes evolve independently of the fast and
that the total energy is composed of both slow and fast components,
while the potential enstrophy is, to leading order in the expansion
parameter, purely slow. The reduced equations and their conservation
laws are supported by numerical simulations using low wave number
forcing. These slow equations are not quasi-geostrophic because
they are nonhydrostatic.

%

\section{Nonlocal form of the Boussinesq equations}
\label{sec:nonlocal}
The Boussinesq equations for flow moving at a constant rotation about
the $\mathbf{\widehat{z}}$ axis for vertically stratified flow is,
\begin{gather}
  \label{eq:bousnddim} 
  \frac{D}{Dt}\mathbf{v} +f ~\hat{\mathbf z}\times \mathbf{v}
  + \rho_0^{-1} \rho g \mathbf{\hat{z}}  + 
  \rho_0^{-1} \grad  p  = \nu \Delta \mathbf{v}, \\    
  \label{eq:buoynddim}  
  \frac{D}{Dt} \rho - b w = \kappa \Delta \rho,\\
  \grad\cdot \mathbf{v} = 0,
\end{gather}
where $\frac{D}{Dt} = \frac{\partial}{\partial t} + \mathbf{v} \cdot
\grad$ is the material derivative, $\mathbf{v}=(u,v,w)$ is the
Eulerian velocity, $p$ is the pressure and the total density 
$\tilde{\rho}$ has been decomposed into 
$\tilde{\rho} = \rho_o - b z + \rho$, where $\rho_0$ is 
a constant background reference value of the density, $b$  is
the density gradient in the vertical, and $\rho$ is the density 
fluctuation.  We assume $b>0$ for stable stratification. The parameter 
$f$ is twice the frame rotation rate, $g$ is the acceleration of gravity, 
$\nu$ is the kinematic viscosity, and $\kappa$ the diffusion coefficient.

In order to distinguish the physical mechanisms of fast rotation from
weak stratification we use the same velocity and length scales for all
three components of velocity and in all three dimensions. Therefore we
nondimensionalize using the following characteristic scales; $L$ is
the length scale for the three spatial coordinates $\mathbf{x} =
(x,y,z)$, $U$ is the velocity scale, and $L/U$ is the advective time
scale. The scale for the density fluctuation is $bU/N$, where $N=(g b/
\rho_o)^{1/2}$ is the Brunt-V\"ais\"al\"a frequency.  Then we arrive
at the following nondimensional quantities,
\begin{gather}
  \label{eq:nondparams}
  Ro = \frac{U}{f L}, \quad Eu = \frac{P}{\rho U^2}, \quad Re =
  \frac{U L}{\nu}, \quad Pr = \frac{\nu}{\kappa},\quad Fr = \frac{U}{N L},
\end{gather}
where $Ro$ is the Rossby number, $Fr$ is the Froude number, $Eu$ is the
Euler number, $Re$ is the Reynolds number, and $Pr$ is the Prandtl number.
Then the non-dimensional Boussinesq equations for rotating and stratified 
flow are, 
\begin{gather}
  \label{eq:bousnd} 
  \frac{D}{Dt}\mathbf{v} +\frac{1}{Ro}\hat{\mathbf z}\times \mathbf{v} +
   Eu~\grad p + \frac{1}{Fr} \rho \mathbf{\hat{z}} = 
  \frac{1}{Re} \Delta \mathbf{v}, \\  
  \label{eq:buoynd} 
  \frac{D}{Dt}\rho  -\frac{1}{Fr}~w= \frac{1}{Re Pr} \Delta \rho  
  \quad \text{with} \quad \grad\cdot \mathbf{v} = 0.
\end{gather}

It is clear that $\mathbf v$ and $\rho$ are the evolution variables 
in Eqs. (\ref{eq:bousnd})-(\ref{eq:buoynd}) and that the role of the 
pressure gradient term in the momentum equation is to enforce the 
incompressibility condition. By eliminating the pressure term it is 
possible to recast the Boussinesq equations exclusively in terms of
the evolution variables and at the same time to incorporate the 
incompressibility constraint. This equivalent formulation is however
in nonlocal form.  To write these equations in their nonlocal form take 
the divergence of the momentum equation to find the equation for the 
pressure,
\begin{gather}
    \label{eq:empresnd}
    Eu \mathbf \grad  p =  \grad \Delta^{-1} \biggl(  
    \frac{1}{Ro} \mathbf{\hat{z}} \cdot {\bm \omega} - \frac{1}{Fr}  
    \frac{\partial \rho}{\partial z} - \grad \cdot (\mathbf v \cdot
    \grad \mathbf v) \biggr), 
\end{gather}
where $\Delta^{-1}$ is the inverse Laplacian operator and
$\bm{\omega} = \mathbf{\grad \times v} = (\xi, \eta, \omega)$ 
is the local vorticity.  Substitute the equation for the pressure into 
Eqs. (\ref{eq:bousnd})-(\ref{eq:buoynd}) to get the nonlocal form of 
the Boussinesq equations,
\begin{eqnarray}
  \label{eq:bousnd2} 
  &&\frac{D}{Dt}\mathbf{v} +\frac{1}{Ro}\hat{\mathbf z}\times
  \mathbf{v} + \grad   \Delta^{-1} \biggl(     
  \frac{1}{Ro} \mathbf{\hat{z}} \cdot \bm{\omega} - \frac{1}{Fr}
  \frac{\partial \rho}{\partial z} - \grad \cdot 
  (\mathbf v \cdot  \grad \mathbf v) \biggr)  
  + \frac{1}{Fr}\rho \mathbf{\hat{z}} = \frac{1}{Re}
  \Delta \mathbf{v}, \\      
  \label{eq:buoynd2} 
  &&\frac{D}{Dt}\rho - \frac{1}{Fr} ~w = \frac{1}{Re Pr} \Delta\rho. 
\end{eqnarray}
These equations automatically incorporate the incompressibility condition.
Indeed, taking the divergence of Eq. (\ref{eq:bousnd2}) results in 
$\frac{\partial}{\partial t} ( \grad \cdot \mathbf v) =
\frac{1}{Re} \Delta  ( \grad \cdot \mathbf v)$, so that if 
$\grad \cdot \mathbf v$ is zero initially, then it remains zero for all time.

A quantity of fundamental importance is the potential vorticity 
$\tilde{q} = {\bm \omega}_a  \cdot \grad \tilde{\rho}$, 
where ${\bm \omega}_a = {\bm \omega} + f \mathbf{\hat{z}}$. 
Clearly $\tilde{q} =  q - fb$, where the evolution of  
$q = f \frac{\partial \rho}{\partial z} - b \omega + 
{\bm \omega} \cdot \grad \rho$ is given by Ertel's theorem,  
\begin{gather}
\frac{Dq}{Dt} = \nu \Delta {\bm \omega} \cdot \grad \rho
+ \kappa \grad ( \Delta \rho ) \cdot {\bm \omega}_a. 
\end{gather}
If we scale $q$ with $fbFr$ and $\bm \omega$ with $fRo$ then the
nondimensional form of $q$ is,
\begin{gather}
\label{eq:potvorndexpand}  
q = \frac{\partial \rho}{\partial z} - \frac{Ro}{Fr} ~\omega 
+ Ro ~({\bm \omega} \cdot \grad \rho) ,
\end{gather} 
and the nondimensional form of Ertel's equation for  $q$ is, 
\begin{equation}
    \label{eq:ndpotvoreqn}
    \frac{Dq}{Dt}   
    = \frac{1}{Re Pr} \Delta \frac{\partial \rho}{\partial z} 
    - \frac{Ro}{Fr Re} \Delta \omega 
    + \frac{Ro}{Re} \Delta {\bm \omega} \cdot \grad \rho
    + \frac{Ro}{Re Pr} {\bm \omega} \cdot \grad \Delta \rho .
\end{equation}
The equations for the global integrated total energy and potential
enstrophy are,
\begin{equation}
  \label{eq:energynd}
  \frac{1}{2} \frac{d}{dt} ~\int_V ~ (|\mathbf{v}|^2 + 
  \rho^2 ) \; d \slv = -\frac{1}{Re} \int_V | \grad \mathbf v|^2 \; d \slv 
  - \frac{1}{Re Pr} \int_V | \grad \rho |^2 \; d \slv,  
\end{equation}
\begin{equation}
  \label{eq:ndpotens}
  \frac{1}{2} \frac{d}{dt} ~\int_V q^2 \; d \slv
   =  \int_V  q \frac{\partial q}{\partial t} \; d \slv =
   - \frac{1}{Re Pr} \int_V ~\biggl| \grad 
   \frac{\partial \rho}{\partial z} \biggr|^2 \; d \slv 
   + O(Ro).
\end{equation}
The energy equation, Eq. (\ref{eq:energynd}) is independent
of the Rossby and Froude number but the enstrophy equation
Eq. (\ref{eq:ndpotens}) depends on the Rossby number, with a leading dissipative term 
depending on  $ |\grad \frac{\partial \rho}{\partial z}|^2$ and the 
remaining contributions  involving powers of the Rossby number.

\section{Limiting dynamics for the rapidly rotating Boussinesq equations}
\label{sec:limiting}
Here we formulate the limiting dynamics for the rapidly rotating
Boussinesq equations, i.e. in the limit of $Ro \rightarrow 0$ and $Fr
= O(1)$. In doing so we will follow the approach developed in in great
generality by Embid and Majda (1998) and which builds upon earlier
work of \cite{KLAINERMANS:SINLOQ}, \cite{MajdaCompBook} and
\cite{schochet1987,schochet1994}. In fact, the present work
complements the work of Embid and Majda which focused on the cases
where $Fr \rightarrow 0$ with either $Ro/Fr$ finite or $Ro = O(1)$.
The analysis starts with the recasting of the rotating Boussinesq
equations in an abstract setting that reveals its key structure.  This
is followed with the asymptotic formulation of the slow dynamics
equations in the limit of $Ro \rightarrow 0$ and balanced initial
data, i.e. without fast inertial waves. Finally, we adapt the theory
developed by Embid and Majda to formulate limiting dynamics equations
in the limit of $Ro \rightarrow 0$ and with fast inertial waves.
\subsection{Abstract framework for the rotating Boussinesq equations}
If we introduce the vector $\bfu = ( \bfv , \rho)$ and let $Ro = \eps$, 
then the rotating Boussinesq equations, Eqs. (\ref{eq:bousnd2}) -
(\ref{eq:buoynd2}) become, in abstract operator form, 
\begin{eqnarray}
\label{eq:operatorbous}
&&\frac{\partial \mathbf{u}}{\partial t} + \frac{1}{\epsilon} L_F
\mathbf{u} +  L_S
\mathbf{u} + B(\mathbf{u},\mathbf{u}) = D\mathbf{u}, \\
&&\mathbf{u}|_{t=0} = \mathbf{u_0}(\mathbf{x}), \nonumber
\end{eqnarray}
with the operators $L_F$, $L_S$, $B$ and $D$ given by,
\begin{eqnarray}
\label{eq:operatorros}
&& L_F \mathbf{u} =
\begin{pmatrix}
\mathbf{\widehat{z} \times v} + \mathbf{\nabla} \Delta^{-1} \omega\\
0
\end{pmatrix}
\\
\label{eq:operatorfr}
&&L_S\mathbf{u}= (Fr)^{-1}
\begin{pmatrix}
 \mathbf{\rho ~\widehat{z}}  - \grad  \Delta^{-1} (
 \frac{\partial \rho}{\partial z})\\
-w
\end{pmatrix}
\\
&&B(\mathbf{u},\mathbf{u}) =
\begin{pmatrix}
\mathbf{v} \cdot \grad \mathbf{v} - \grad
\Delta^{-1} (\grad \cdot (\mathbf{v} \cdot \grad \mathbf{v}))\\
\mathbf{v} \cdot \grad \rho
\end{pmatrix}
\\
\label{eq:operatorD}
&&D\mathbf{u} =
\begin{pmatrix}
(Re)^{-1} \Delta \mathbf{v}\\
(Re)^{-1} (Pr)^{-1} \Delta \rho
\end{pmatrix} .
\end{eqnarray}
In the equations above the linear operator $L = \eps^{-1}L_F + L_S$ splits
into a fast piece $L_F$ associated with the Rossby number $Ro = \eps$ and
a slow piece $L_S$ associated with the Froude number. It is clear that
Eq. (\ref{eq:operatorbous}) becomes singular in the limit of 
$\eps \rightarrow 0$, and the fast operator $L_F$  will have a dominant 
role. The remaining terms in Eq. (\ref{eq:operatorbous}) 
are given by the bilinear advective operator $B(\bfu, \bfu)$ and the 
diffusion operator $D\bfu$. 

As we mentioned before,
if the initial data $\bfu_0$ in Eq. (\ref{eq:operatorbous}) is divergence
free, then the solution remains divergence free for all time. But in fact
more is true, each individual operator $L_F$, $L_S$, $B$ and $D$ takes 
solenoidal fields into solenoidal fields. 
Therefore a natural setting for Eq. (\ref{eq:operatorbous}) is the Hilbert
space $X$ of vector fields $\bfu = (\bfv , \rho)$ in $L^2$ that are 
divergence free, $\grad \cdot \bfv = 0 $, and equipped with the 
$L^2$ -- norm, which is physically equivalent to the total energy, 
$\| \bfu \|^2 = \int | \bfv |^2 + \rho^2 \; d \slv$. In addition, we assume 
$2\pi$-periodicity in all the space variables. This choice of boundary conditions
considerably simplifies the study of Eq. (\ref{eq:operatorbous}), particularly   
the analysis of the operator $L_F$, and the resulting slow limiting dynamics 
equations, Eq. (\ref{eq:slowdyn}), and the fast wave averaging equations, Eq. (\ref{eq:fastdyn}).
The reason for this simplification is the fact that the associated eigenfunctions are 
given explicitly in terms of Fourier modes. In addition, the choice of periodic boundary
conditions is consistent with the numerical simulations presented in Section 4. 
Changing the domain and the boundary conditions can make the mathematical analysis quite 
difficult; for example, for arbitrary bounded domains it may be impossible to characterize
the eigenfunctions of $L_F$. The choice of an infinite domain may change drastically the
structure of the null space (slow waves) and the range (fast waves) of the operator
$L_F$.

One key observation is the fact that the operator $L_F$ (and
$L_S$) is skew-Hermitian in $X$: for $\bfu_1$ and $\bfu_2$ in $X$,
\begin{equation}
\int_V \bfu_2^* L_F \bfu_1 \; d \slv = - 
\int_V  ( L_F \bfu_2)^* \bfu_1 \; d \slv , 
\end{equation} 
where $\bfu^*$ denotes the conjugate transpose of $\bfu$.
Several important properties follow from this fact. First, 
Eq. (\ref{eq:operatorbous}) satisfies (in the absence of diffusion) the
conservation of energy, Eq. (\ref{eq:energynd}) . This property is 
shared with other important systems in mathematical physics, such as 
the Euler and the Maxwell equations. Second, according to the Spectral
Theorem, skew-Hermitian operators have purely imaginary eigenvalues and 
an orthonormal basis of eigenfunctions, see \cite{Lax}. Physically this 
means that the basic normal mode solutions of equations represent  
wave motions. Finally, the null space of
$L_F$, $N(L_F)$, is orthogonal to the range of $L_F$, $R(L_F)$. This can
be thought of a consequence of the Spectral Theorem because $N(L_F)$
is spanned by the eigenfunctions with zero eigenvalue (slow modes) whereas 
$R(L_F)$ is spanned by the remaining eigenfunctions with non-zero 
eigenvalues (fast modes). This last property will be exploited later in the derivation 
of the slow dynamics equations.

Next we apply the previous observations to the linear equation
\begin{equation}
\label{eq:linearfast}
\frac{\partial \mathbf{u}}{\partial t} +
\frac{1}{\eps} L_F \mathbf{u} = 0,
\end{equation}
and seek normal mode solutions in the form of harmonic plane waves 
\begin{equation}
\mathbf{u} (\bfx, t) =  \bfr \exp  \left[ i \mathbf{k} \cdot \mathbf{x} - i
\eps ^{-1} \omega(\mathbf{k}) ~t \right],
\end{equation}
where $\bfk = (k,l,m)$ is the wave number, $\omega (\bfk)$ is the 
frequency and the purely imaginary number $\lambda = i \omega (\bfk )$ is 
the eigenvalue of $L_F$ associated with the eigenfunction 
$\bfu = \bfr \exp \left[ i \bfk \cdot \bfx \right]$. The four eigenfrequencies 
$\omega ( \bfk )$ are given by the dispersion relations
\begin{eqnarray}
  \label{eq:freqr0}
  \omega(\mathbf{k}) =\pm ~m / | \bfk |, \quad \omega(\mathbf{k}) = 0
  \quad \text{(double).}
\end{eqnarray}
Therefore the equations admit slow modes moving on time scales
$O(1)$ when $\omega ( \bfk ) = 0$ and fast waves moving on time scales 
$O(1/\epsilon)$ when $\omega ( \bfk ) \neq  0$. 
The fast waves in this limit are {\bf{gyroscopic or inertial
waves}}. They are waves who owe their existence to the presence of
the Coriolis force and were originally described by \cite{Kelvin1880}.
Descriptions of these waves can be found in \cite{LeBlondMysak} and
\cite{Greenspan}. Of course, if $m=0$ in Eq. (\ref{eq:freqr0}) then we only have
slow gyroscopic waves.
Explicit formulas for the eigenvectors $\bfr$ associated
with the fast and slow normal modes are given in the appendix.
\subsection{Slow limiting dynamics as $Ro \rightarrow 0$}
Here we consider the limiting dynamics equations as $Ro \rightarrow 0$
under the assumption that the solution $\mathbf{u}^\eps ( \mathbf{x} , t)$  
of Eq. (\ref{eq:operatorbous}) evolves only on the slow (advective) time scale. 
The formal derivation in the context of the abstract operator equation, 
Eq. (\ref{eq:operatorbous}), is straightforward. 
We start by assuming that $\bfu^\eps ( \bfx, t)$  has the asymptotic expansion 
\begin{equation}
\bfu^\eps (\bfx, t)= \bfu^0(\bfx, t) + \eps \bfu^1(\bfx, t) + O(\eps^2),
\end{equation}
as $\eps \rightarrow 0$. Plugging $\bfu^\eps$ into 
Eq. (\ref{eq:operatorbous}) and collecting the contribution of order 
$O(\epsilon^{-1})$ yields
\begin{equation}
\label{eq:orderzero}
L_F \mathbf{u}^0 \equiv 0,
\end{equation}
that is, $\mathbf{u}^0$ is in $N(L_F)$ for all time, or equivalently, $\bfu^0 (\bfx, t)$ 
is represented exclusively in terms of slow modes. In particular, the
initial data $\bfu_0 (\bfx) = \bfu^0 (\bfx, 0)$ is in $N(L_F)$ to 
leading order in $\eps$. The next  contribution of order 
$O(\epsilon^0)$ yields
\begin{equation}
\label{eq:orderone}
\frac{\partial \mathbf{u}^0}{\partial t} + L_F \mathbf{u}^1 +
L_S \mathbf{u}^0 + B(\mathbf{u}^0, \mathbf{u}^0) - D\mathbf{u}^0 = 0.
\end{equation}
The slow limiting dynamics equation is now obtained by projecting 
Eq. (\ref{eq:orderone}) onto $N(L_F)$ as follows.
First apply the orthogonal projection $P$ of $X$ onto $N(L_F)$ to both sides
of Eq. (\ref{eq:orderone}). Since $\bfu^0$ is in $N(L_F)$ for all time so is
$\partial \bfu^0 / \partial t$, hence $P( \partial \bfu^0 / \partial t ) = 
\partial \bfu^0 / \partial t $. In addition, since $L_F \bfu^1$ is in 
$R(L_F)$, and $N(L_F)$ is orthogonal to $R(L_F)$, then $P( L_F \bfu^1) = 0$
and any contribution from  $\bfu^1$ is eliminated  under the projection. 
Finally, we eliminate the superscript in $\bfu^0$ and obtain the slow 
limiting dynamics equations
\begin{eqnarray}
\label{eq:slowdyn}
&& \frac{\partial \bfu }{\partial t} + P( L_S \bfu  + 
B( \bfu , \bfu ) - D\bfu ) = 0, \\
&&\bfu|_{t=0} =  \bfu_0(\mathbf{x}) \in N(L_F).  \nonumber
\end{eqnarray}
Next we notice that, to leading order in $\eps$, it is enough for the
initial data $\bfu_0(\bfx)$ to be in $N(L_F)$ to automatically guarantee
that the solution $\bfu(\bfx, t)$ of Eq. (\ref{eq:slowdyn}) remains 
in $N(L_F)$ for all time.
Indeed, if we integrate in time  Eq. (\ref{eq:slowdyn}) and use the fact
that $\bfu_0 \in N(L_F)$, we conclude that $\bfu(t) \in N(L_F)$ for all time.
Moreover, we will show shortly that in the  context of the rotating 
Boussinesq equations the null space $N(L_F)$ consists precisely of the 
{\bf Taylor-Proudman} columnar flows, \cite{Taylor}. 
Therefore we can say that to leading order in $\eps$, if the initial data 
is a Taylor-Proudman column (i.e. free of fast inertial waves), then the
solution $\bfu$ remains a Taylor-Proudman column for all time and its 
evolution is described by the slow dynamics equations. When the initial data
is a Taylor-Proudman column to leading order in $\eps$ we say that the flow is
in approximate Taylor-Proudman balance. Finally we remark that all these 
formal considerations can be established with full mathematical rigor 
through a direct application of the general theory of singular limits of 
hyperbolic systems first developed by  \cite{KLAINERMANS:SINLOQ},  
\cite{MajdaCompBook}, with later additions by \cite{schochet1987}.

To obtain the concrete formulation of the slow dynamics for the 
Boussinesq equations, Eqs. (\ref{eq:operatorbous})- (\ref{eq:operatorD}),
we need to determine explicitly the null space $N(L_F)$ and its
orthogonal projection $P$. For this purpose it is convenient to
split vectors and operators into their horizontal and vertical components.
Thus, the velocity $\bfv = (\bfv_H , w)$ with $\bfv_H = (u, v)$, the gradient
$\grad = (\gradH, \dd{}{z})$ with $\gradH = (\dd{}{x} , \dd{}{y})$, and
the Laplacian $\Delta = \Delta_H + \dd{^2}{z^2}$ with 
$\Delta_H = \dd{^2}{x^2} + \dd{^2}{y^2}$. 

The null space $N(L_F)$ of the fast operator $L_F$ in 
Eq. (\ref{eq:operatorros})  is characterized by, 
\begin{eqnarray}
\label{eq:nullx}
-v + \dd{}{x} \Delta^{-1} \omega  &=& 0,\\
\label{eq:nully}
u + \dd{}{y} \Delta^{-1} \omega  &=& 0,\\
\label{eq:nullz}
\dd{}{z} \Delta^{-1} \omega &=& 0,
\end{eqnarray}
where $\omega = \dd{v}{x} - \dd{u}{y}$  is the vertical component
of the vorticity and $\bfv$ is incompressible,  $\grad \cdot \bfv = 0$.
From Eq. (\ref{eq:nullz}) it follows that $\Delta^{-1} \omega = \psi$
is $z$--independent of $z$. Introducing $\psi$ back into  
Eqs. (\ref{eq:nullx})--(\ref{eq:nully}), shows that $\psi$ is the 
streamfunction for $\bfv_H$, $\bfv_H = ( -\dd{\psi}{y}, \dd{\psi}{x})$,
and that $\bfv_H$ is incompressible, $\gradH \cdot \bfv_H = 
\dd{}{x}(-\dd{\psi}{y}) + \dd{}{y}(\dd{\psi}{x}) = 0$. Since $\bfv$ is 
incompressible by assumption, then it follows that $\dd{w}{z}=0$, i.e. 
$w$ is also $z$--independent. This shows that $N(L_F)$ consists of 
Taylor-Proudman column flows, i.e. states $\bfu = (\bfv, \rho)$ with 
$\bfv$ $z$--independent and $\bfv_H$ incompressible. That no restrictions
are imposed upon $\rho$ is not surprising since the fast operator $L_F$
only includes those contributions associated with the Rossby number. 

The orthogonal projection $P$ onto the null space $N(L_F)$ is given by,
\begin{eqnarray}
\label{eq:projontonull}
P\mathbf{u} =
\begin{pmatrix}
\avez{\bfv_H} - \grad_H \Delta_H^{-1} ( \grad_H \cdot \avez{\bfv_H}) \\
\avez{w} \\
\rho
\end{pmatrix},
\end{eqnarray}
where $\avez{f} = \frac{1}{2\pi} \int_0^{2\pi} f(x,y,z) \; dz$ denotes 
the average in the vertical direction. Therefore the concrete form 
of Eq. (\ref{eq:slowdyn}), the {\bf slow limiting dynamics equations} for the 
rotating Boussinesq equations is,  
\begin{eqnarray}
\label{eq:2dns}
&&\frac{\partial \mathbf{v}_H}{\partial t} + \mathbf{v}_H \cdot
  \gradH \mathbf{v}_H + \gradH p = \frac{1}{Re}
  \Delta_H \mathbf{v}_H\\
\label{eq:2div}
&&\gradH \cdot \bfv_H = 0 \\
\label{eq:2dw}
&&\frac{\partial w}{\partial t} + \mathbf{v}_H \cdot
 \gradH  w = \frac{1}{Re} \Delta_H w - \frac{1}{Fr}  \avez{\rho} \\
\label{eq:3drho}
&&\dd{\rho}{t} + \bfv \cdot \grad \rho  - \frac{1}{Fr} w = 
\frac{1}{Re Pr}  \Delta \rho , 
\end{eqnarray}
where $\bfv = \bfv(x,y,t)$, $\rho = \rho(x,y,z,t)$, and $\gradH p =
\gradH \Delta_H^{-1} [ \gradH \cdot (\bfv_H \cdot \gradH \bfv_H)]$.
For brevity of notation we omit distinguishing between projected and
unprojected variables and instead state that all the variables in
Eqs. (\ref{eq:2dns})-(\ref{eq:3drho}) are the result of applying the
projection operator, Eq (\ref{eq:projontonull}).

In the slow dynamics the horizontal component of the velocity 
$\bfv_H$ is governed by the 2D Navier-Stokes equation. Moreover, $\bfv_H$  evolves 
independently of the vertical velocity $w$ and the density $\rho$ but it influences
the dynamics of these variables through the advection terms in 
Eqs. (\ref{eq:2dw}) -- (\ref{eq:3drho}). The dynamics of the 
vertical velocity $w$ and the density $\rho$ are strongly coupled. Interestingly,
the vertical velocity $w$ evolves according to a 2D forced advection-diffusion
equation, Eq. (\ref{eq:2dw}), with buoyancy forcing given by $\avez{\rho}$, 
the density average in the vertical direction. On the other hand, the evolution 
of the density $\rho$  is given by the 3D forced advection-diffusion equation,
Eq. (\ref{eq:3drho}), and remains the same as Eq. (\ref{eq:buoynd}) in the Boussinesq 
approximation. 

A consequence of this decoupling of the horizontal velocity from  the vertical 
velocity and the density in Eqs. (\ref{eq:2dns}) -- (\ref{eq:3drho}) is the 
appearance of additional globally integrated conservation laws which are not
present in the original Boussinesq equations.  First, the 2D Navier-Stokes equation 
for $\bfv_H$  yields the conservation of horizontal kinetic energy and 
enstrophy,
\begin{eqnarray}
\label{eq:2dhke}
&& \frac{1}{2} \frac{d}{d t} \int_A ~|\bfv_H|^2 \; d \sla  =
- \frac{1}{Re} \int_A | \gradH \bfv_H |^2 \; d \sla ,  \\
\label{eq:2denstrophy}
&&\frac{1}{2} \frac{d}{d t} \int_A \omega^2 \; d \sla =
-\frac{1}{Re} \int_A | \gradH \omega |^2 \; d \sla ,
\end{eqnarray}
where integration is over the horizontal period square $A = [0, 2\pi]^2$.
Next, taking the vertical average of the equation for $\rho$, Eq. (\ref{eq:3drho}),
results in an evolution equation for $\avez{\rho}$, 
\begin{equation}
\label{eq:2drhoz}
\dd{}{t} \avez{\rho} + \bfv_H \cdot \gradH \avez{\rho}
- \frac{1}{Fr} w = \frac{1}{Re Pr} \Delta_H \avez{\rho}.
\end{equation}
Combining Eq. (\ref{eq:2dw}) and Eq. (\ref{eq:2drhoz}) results in horizontal
conservation laws for the vertical kinetic energy and the average vertical
potential energy,  
\begin{eqnarray}
\label{eq:2dvke}
&& \frac{1}{2} \frac{d}{d t} \int_A w^2 \; d \sla  =
- \frac{1}{Fr} \int_A w \avez{\rho} \; d \sla
- \frac{1}{Re} \int_A | \gradH \; w |^2 \; d \sla ,  \\
\label{eq:2dpe}
&& \frac{1}{2} \frac{d}{d t} \int_A  \avez{\rho}^2 \; d \sla =
\frac{1}{Fr} \int_A w \avez{\rho} \; d \sla
- \frac{1}{Re Pr} \int_A | \gradH \avez{\rho}|^2 \; d \sla , 
\end{eqnarray}
and adding these two equations results in the conservation of vertical energy,
\begin{equation}
\label{eq:2dve}
\frac{1}{2} \frac{d}{d t} \int_A ( w^2 +   \avez{\rho}^2) \; d \sla =
-\frac{1}{Re} \int_A | \gradH \; w |^2 \; d \sla
- \frac{1}{Re Pr} \int_A | \gradH \avez{\rho}|^2 \; d \sla.
\end{equation}
Additionally, if we define the density fluctuation $\rhot = \rho - \avez{\rho}$,
then from the density equations, Eqs. (\ref{eq:3drho}) -- (\ref{eq:2drhoz}), it 
follows the conservation of potential energy,
\begin{equation}
\label{eq:3dpe}
\frac{1}{2} \frac{d}{d t} \int_V  \rhot^2 \; d \slv =
- \frac{1}{Re Pr} \int_V | \grad \rhot|^2 \; d \slv , 
\end{equation}
where the volume integral is over the period cube $V = [0, 2\pi]^3$.
Of course, Eq. (\ref{eq:energynd}) for the conservation of total energy still 
holds for the slow  dynamics equations, Eqs. (\ref{eq:2dns}) -- (\ref{eq:3drho}). 
Finally, define the potential vorticity $q$ for the slow  dynamics equations
as the the leading term of the expansion of $q$ given by Eq. 
(\ref{eq:potvorndexpand}) in powers of $Ro = \eps$, 
$q = \dd{\rho}{z} = \dd{\rhot}{z}$. Then the equation for the conservation 
of potential enstrophy is 
\begin{equation}
\label{eq:consvpotenslow}
\frac{1}{2} \frac{d}{d t} \int_V q^2 \; d \slv =
- \frac{1}{Re Pr} \int_V | \grad q |^2 \; d \slv.
\end{equation}

The conservation laws above were obtained directly from the slow  limiting dynamics 
equations. In the different limiting regimes of strong stratification and Burger number 
$Bu = (Ro/Fr)^2 =  0(1)$, or strong stratification and weak rotation, \cite{BabinA:Onara, Babin1998}
showed that conserved quantities for the limiting equations, like the horizontally 
averaged buoyancy, correspond to adiabatic invariants for the full Boussinesq equations. 

Finally, we remark that although the abstract derivation of the slow limiting dynamics equations, 
Eq. (\ref{eq:slowdyn}), is completely general, the concrete form of the equations depend on the
explicit calculation of $N(L_F)$, $R(L_F)$ and the projection operator $P$ of $X$ onto $N(L_F)$.
This calculation is very much dependent on the choice of the domain and the boundary conditions.
For example, if we assume the horizontal variables infinite in extent and periodicity in the
vertical variable, i.e. $V = \RR^2 \times [0, 2\pi]$, then $N(L_F)$ is still given by Taylor-Proudman
columns, Eqs. (\ref{eq:nullx}) -- (\ref{eq:nullz}), the projection operator by Eq. (\ref{eq:projontonull}),  and the slow limiting 
dynamics equations by Eqs. (\ref{eq:2dns}) -- (\ref{eq:3drho}). On the other hand, if we also assume infinite extent in 
the vertical variable, i.e. $V = \RR^3$, then $N(L_F)$ only admits flows with  zero velocity (this is 
reasonable because the only velocity field $\bfv$ that is $z$-independent and has finite kinetic energy
is $\bfv = \bf0$) and in this case the slow limiting dynamics becomes trivial. However, it is arguable whether
a fluid of infinite depth constitutes a reasonable assumption in the present context.

\subsection{Fast limiting dynamics as $Ro \rightarrow 0$ } 
\label{subsec:fastlimit}
The derivation of the slow limiting dynamics equations presented in the previous 
section was based on the assumption that the solution evolved only on 
the slow advective time scale. This assumption is warranted if, to leading order in $\eps$,
the initial data does not include fast inertial waves components.  If on the other hand  
the initial data  contains inertial waves, then the limiting dynamics equations as 
$Ro \rightarrow 0$ must be modified to take into account the fast inertial waves.
The derivation of the fast limiting dynamics for small Rossby number and finite
Froude number is readily obtained by invoking the  very general approach developed in
the fundamental work of Embid and Majda (1998) on the limiting dynamics of
the Boussinesq equations with small Froude number and either finite or small Rossby 
number.  In fact, we only have to switch the roles of the fast and slow operators 
$L_F$ and $L_S$ and apply their theory in straightforward fashion. For this reason
we are content with a summary of the main points of their theory and their most 
relevant  conclusions for the present work. 

Following \cite{EmbidPF:LowFnl} we assume that the solution 
$\bfu^\eps(\bfx,t)$ of Eq. (\ref{eq:operatorbous}) depends on two separate 
time scales, the slow advective time scale $t$ and the fast time scale 
$\tau = t/\eps$ associated with the inertial waves. 
In addition, we assume that for $\eps \ll 1$ the solution has the 
asymptotic expansion,
\begin{equation}
\label{eq:multscalexpan}
\mathbf{u}^\epsilon(\mathbf{x},t) =
\mathbf{u}^0(\mathbf{x},t,\tau)|_{\tau = t/\epsilon} + \epsilon
\mathbf{u}^1(\mathbf{x},t,\tau)|_{\tau = t/\epsilon} +O(\eps^2) ,
\end{equation}
and it is also assumed that $\bfu^1 ( \bfx, t, \tau) = o(\tau)$, uniformly 
on $0 \leq \tau \leq T/\eps$, to guarantee the asymptotic validity of the
expansion. The analysis of Embid and Majda then shows that to leading order in
$\eps$ the solution $\bfu^\eps(\bfx,t)$ of Eq. (\ref{eq:operatorbous}) is given by,
\begin{equation}
\label{eq:lowestordsoln}
\bfu^\eps (\bfx,t) = \bfu^0 (\bfx,t,\tau)|_{\tau = t/\eps}  + o(1) = 
e^{-t/\eps L_F} \bfu(\bfx,t) + o(1), 
\end{equation}
where $\bfu(\bfx,t)$ solves a reduced equation obtained by averaging
over the fast time variable $\tau$,
\begin{eqnarray}
\label{eq:fastdyn}
&& \dd{\bfu}{t}  + 
   \lim_{T\rightarrow\infty} \frac{1}{T} \int_0^T
   e^{\tau L_F} \left[
   L_S ( e^{-\tau L_F}\bfu) + 
   B(e^{-\tau L_F} \bfu ,e^{-\tau L_F} \bfu ) 
  -D(e^{-\tau L_F} \bfu) \right] d \tau = 0, \ \  \\ \nonumber
&& \bfu (\bfx,t)|_{t=0} = \bfu_0(\bfx). 
\end{eqnarray}
The fast wave averaging equation, Eq. (\ref{eq:fastdyn}), supersedes the slow 
dynamics equation, Eq. (\ref{eq:slowdyn}), whenever inertial waves are present.
We also remark that the asymptotic analysis of \cite{EmbidPF:LowFnl}
can be justified with complete mathematical rigor via the technique of 
cancellation of oscillations developed in the important paper of \cite{schochet1994}.

In practice it may be difficult to evaluate the limit over the fast variable
$\tau$ in Eq. (\ref{eq:fastdyn}). However, this calculation can be performed in
the case of periodic boundary conditions. In this case the fast operator $L_F$ 
has an orthonormal basis of periodic eigenfunctions of the form 
\begin{equation}
\label{eq:eigfun}
\bfu_\bfk^\za (\bfx) = e^{i \bfk \cdot \bfx} \bfr^\za_\bfk,
\end{equation}
where $\bfk = (k,l,m)$ is the wave number,  $\za$ indicates whether the mode 
is slow ($\za = 0$) or fast ($\za = \pm 1$), and 
$\bfr_\bfk^\za = (\bfv_\bfk^\za, \rho_\bfk^\za)$. The associated  purely 
imaginary eigenvalue is $\lambda_\bfk^\za = i \w_\bfk^\za$, where the frequency
$\w_\bfk^\za$ is given by Eq. (\ref{eq:freqr0}), namely 
$\w_\bfk^{\pm 1} = \pm m/|\bfk| $ and $\w_\bfk^0 = 0$.
The explicit form of these eigenfunctions is given in the Appendix.  
Next, we expand $\bfu(\bfx,t)$ in terms of the eigenfunctions of $L_F$,    
\begin{equation}
  \label{eq:eigenexpansion}
  \bfu (\bfx,t) = \sum_{\mathbf{k}}
  \sum_{\alpha=-1}^{1} \sigma_{\mathbf{k}} ^{\alpha}(t) 
    e^{i \mathbf{k} \cdot \mathbf{x}} \mathbf{r}_{\mathbf{k}}^{\alpha},
\end{equation}
and introduce this expansion into the fast wave averaging equation,
Eq. (\ref{eq:fastdyn}). In order to evaluate the fast time averaging in 
Eq. (\ref{eq:fastdyn}) we observe that 
\begin{equation}
\label{eq:exptau}
e^{\tau L_F}(e^{i\bfk \cdot \bfx} \bfr_\bfk^\za ) =
e^{i \tau \w_\bfk^\za} e^{i\bfk \cdot \bfx} \bfr_\bfk^\za,
\end{equation}
and also that 
\begin{equation}
\lim_{T \rightarrow \infty} \frac{1}{T} \int_0^T
e^{i \w \tau } \; d\tau = \left\{ 
\begin{matrix}
      1 \mbox{ if } \w = 0 \\
      0 \mbox{ if } \w \neq 0
\end{matrix}
\right. .
\end{equation}
With these observations we can evaluate all the terms in the limiting fast 
dynamics equation, Eq. (\ref{eq:fastdyn}), and conclude that the Fourier 
amplitudes $\sigma^\za_\bfk (t)$ satisfy the system of differential equations 
\begin{equation}
   \label{eq:fastcoef}
   \frac{d \sigma_\bfk^\za}{dt} + 
   \sum_{R_\bfk^\za}  B^{(\za',\za'',\za)}_{(\bfk',\bfk'',\bfk)} 
   \sigma_{\bfk'}^{\za'} \sigma_{\bfk''}^{\za''}  +
   \sum_{S_\bfk^\za} ~L_\bfk^{(\za',\za)}  \sigma_\bfk^{\za'} =
   \sum_{S_\bfk^\za} ~D_\bfk^{(\za',\za)} \sigma_\bfk^{\za'}.
\end{equation}
The first sum in Eq. (\ref{eq:fastcoef}) comes from averaging the nonlinear
advection term $B(\bfu , \bfu)$ with summation over the set of three--wave 
resonant interactions
$R_\bfk^\za =\{ (\bfk', \bfk'', \za', \za'') | 
\bfk' + \bfk'' = \bfk, \w_{\bfk'}^{\za'} + \w_{\bfk''}^{\za''} = \w_\bfk^\za \}$.
The second and third sums come from averaging  the slow operator $L_S$ and the 
diffusion operator $D$ respectively, with summation over the set 
$S_\bfk^\za = \{ \za' | \w_\bfk^{\za'} = \w_\bfk^\za \}$. Formulas for the 
interaction coefficients  $B^{(\za',\za'',\za)}_{(\bfk',\bfk'',\bfk)}$,
$L_\bfk^{(\za',\za)}$ and $D_\bfk^{(\za',\za)}$ are included in the Appendix.

The fast dynamics equations for the Fourier amplitudes in Eq. (\ref{eq:fastcoef})
suggest that there is strong interaction of the fast and slow modes through
three-waves interactions, via the quadratic interaction coefficients
$B^{(\za',\za'',\za)}_{(\bfk',\bfk'',\bfk)}$. However, it is remarkable that 
the dynamics of the slow modes ($\za = 0$) proceeds independently of the fast
modes ($\za = \pm 1$), making the system of slow and fast modes only weakly 
coupled. This is because all the interaction coefficients
$B^{(\pm 1, \pm 1,0)}_{(\bfk',\bfk'',\bfk)}$ corresponding to  
``fast + fast $\rightarrow$ slow'' interaction are always zero. The verification of 
this fact is given in the Appendix. Of course, the equation for the dynamics of
the slow modes in Eq. (\ref{eq:fastcoef}) is nothing more than previously 
derived  slow limiting dynamics equations, Eq. (\ref{eq:slowdyn}), recast in terms
of the Fourier modes. Nevertheless, the slow modes influence the dynamics of the
fast modes in Eq. (\ref{eq:fastcoef}) through ``fast + slow $\rightarrow$ fast ''
interactions. These conclusions mirror those previously derived by  Embid and 
Majda (1998) for the case of small Froude number.

A remarkable consequence of this weak coupling is that in the absence of 
dissipation there is conservation of energy for the slow and the fast modes 
separately. The reasoning, which was first provided by Embid and Majda (1998) 
for the case of small Froude number, is reproduced below. 
First we observe that the solution $\bfu (\bfx ,t)$  of the
fast limiting dynamics, Eq. (\ref{eq:fastdyn}), has a unique orthogonal decomposition 
in terms of slow and a fast component,
\begin{equation}
\mathbf{u}(\mathbf{x},t) =
\mathbf{u}^S(\mathbf{x},t) + \mathbf{u}^F(\mathbf{x},t).
\end{equation}
In fact, $\bfu^S(\bfx, t)$ is given explicitly by Eq. (\ref{eq:eigenexpansion})
with $\za = 0$ and $\bfu^F(\bfx,t)$ by Eq. (\ref{eq:eigenexpansion}) with
$\za = \pm 1$. Next we substitute $\bfu(\bfx,t)$ into 
Eq. (\ref{eq:lowestordsoln}) and make use of Eq. (\ref{eq:exptau}) to conclude
that to leading order in $\eps$ the solution $\bfu^\eps (\bfx,t)$ has the form
\begin{equation}
\bfu^\eps (\bfx,t) = \bfu^S(\bfx,t) + e^{-t/\eps L_F} \bfu^F(\bfx,t) + o(1).
\end{equation}
Moreover, since $L_F$ is a skew-Hermitian operator, then $e^{t/\eps L_F}$ is 
an unitary operator, and since the eigenfunctions of $L_F$ are also an
orthonormal basis for $e^{t/\eps L_F}$ (see Lax (2002)), we conclude that
\begin{equation}
\label{eq:fastenergy}
\| \bfu \|^2 = \| \bfu^S \|^2 + \| e^{t/\eps L_F} \bfu^F\|^2 = 
\| \bfu^S \|^2 + \| \bfu^F\|^2,
\end{equation}
where $\| \bfu \|^2 = \int |\bfv|^2 + \rho^2 \; d \slv$  is twice the total energy
(kinetic plus potential). In the absence of dissipation Eq. (\ref{eq:energynd}) 
shows that the Boussinesq equations conserve energy, so that $\| \bfu \|^2$ is 
constant in time. On the other hand,  the slow limiting dynamics equations also 
conserve energy, so that $\| \bfu^S \|^2$ is also constant in time. 
Combining this two facts with Eq. (\ref{eq:fastenergy}), we conclude 
that $\| \bfu^F\|^2$  is constant in time, thus proving that the  energies
of $\bfu^S$ and $\bfu^F$ are constant separately. This important physical 
property of the fast limiting dynamics equations will be exploited later as 
a diagnostic tool in the numerical simulations. 

Finally, we remark that the Fourier basis in Eq. (\ref{eq:eigfun}) can be used 
to study the dependence of other related physical quantities on the slow and 
fast modes. For example, the leading term of the  potential vorticity $q$ in
Eq. (\ref{eq:potvorndexpand}) has the eigenfunction expansion
\begin{equation}
q = \dd{\rho}{z} = i \sum_\bfk \sum_{\za=-1}^{1}
     m \rho_\bfk^\za \sigma_\bfk^\za (t) ~e^{i \mathbf{k} \cdot \mathbf{x}},
\end{equation}
where $\rho_\bfk^\za$ is the fourth component of the vector $\bfr_\bfk^\za$ 
in Eq. (\ref{eq:eigfun}). Inspection of these vectors in 
Eqs. (\ref{eq:eigv1}) -- (\ref{eq:eigv2}) in the Appendix reveals that
$\rho_\bfk^{\pm 1} = 0$ and in consequence the $q$ is composed of slow modes
even with the presence of fast inertial waves in the fast limiting 
dynamics. This fact can be used as the starting point to prove conservation 
of potential vorticity, in the weak sense and without dissipation, along the
same lines developed in Embid and Majda (1998). By contrast, the the eigenfunction
expansion of the vertical component of the vorticity $\w$ is, 
\begin{equation}
\w = \dd{v}{x} - \dd{u}{y}  = i \sum_\bfk \sum_{\za=-1}^{1}
    (k v_\bfk^\za - l u_\bfk^\za) \sigma_\bfk^\za (t)  
     ~e^{i \mathbf{k} \cdot \mathbf{x}},
\end{equation}
with $u_\bfk^\za$ and $v_\bfk^\za$ being the first and second components of 
$\bfr_\bfk^\za$ in Eq. (\ref{eq:eigfun}). Another inspection of 
Eqs. (\ref{eq:eigv1}) -- (\ref{eq:eigv2}) in the Appendix reveals that
$k v_\bfk^0 - l u_\bfk^0 = 0$ and we conclude that $\w$ is composed exclusively
of fast modes. 

\section{Numerical simulations}
\label{sec:numeric}

\begin{table}
\begin{center}
\begin{tabular} { cccccccccc }
Run Number & $Ro$ & $Fr$ & $f$ & $N$ & $k_f$ & $k_d$ & $\epsilon_f$ & $k_\text{total}$ & $T_0$ \\ 
\hline
1 & 1.0 & 1.0 & 7.0827 & 7.0827 & 3 & 3 & 1.0 & 256 & 20. \\ 
\hline
2 & 0.3 & 1.0 & 23.6091 &  7.0827 & 3 & 10.0 & 1.0 & 256  & 20. \\ 
\hline
3 & 0.2 & 1.0 & 35.4136 &  7.0827 & 3 & 15.0 & 1.0 & 256  & 20.\\ 
\hline
4 & 0.1 & 1.0 & 70.8273 &  7.0827 & 3 & 30.0 & 1.0 & 256  & 20. \\ 
\hline
5 & 0.08 & 1.0 & 88.5341 &  7.0827 & 3 & 37.5 & 1.0 & 256  & 20. \\ 
\hline
6 & 0.05 & 1.0 & 141.6546 &  7.0827 & 3 & 60.0 & 1.0 & 256  & 20. \\
\hline
7 & 0.01 & 1.0 & 708.2731 &  7.0827 & 3 & 300.0 & 1.0 & 256  & 60. \\
\end{tabular}
\caption{This figure tabulates the parameters used in the simulations
  at $256^3$ with low wave number forcing.}
\label{table:Ro256}
\end{center}
\end{table}
The goal of this section is to see if key attributes of the $Fr \approx
1, Ro \rightarrow 0$ limiting dynamics can be reproduced in numerical
simulations that use low wave number white noise forcing. The three
aspects we examine are: 1) the columnar structure, 2)
the time evolution of the ratio of the $E_{Ro \rightarrow 0}$ (slow)
total energy to the total energy, $E$, 3) the time evolution of the
ratio of the $Q_{Ro \rightarrow 0}$ (slow) potential enstrophy to the
total potential enstrophy, $Q$.

For all our simulations, detailed in Table (\ref{table:Ro256}), we use
the triply periodic, pseudo-spectral LANL/Sandia Direct Numerical
Simulation (DNS) code that solves Eq
(\ref{eq:bousnddim})-(\ref{eq:buoynddim}) in a rectangular domains (2D
$[0,1]^2$ or 3D $[0,1]^3$) with a pseudo-spectral method or 4th order
finite differences and a RK4 time stepping scheme.  The code allows
for an arbitrary number of passive scalars and arbitrary aspect ratio
grids.  The code uses MPI for its parallelization along with a 3D
domain decomposition (allowing for slab decomposition, pencil
decomposition or cube decomposition).  Since its inception it has been
designed for performance on massively parallel computers and has
excellent scalability, which has been demonstrated on up to 18,000
processors running problems as large as $4096^3$ (64B grid points).
All diagnostics and associated I/O are also implemented with fully
scalable algorithms.  The parallel FFT at the core of the model is one
of the fastest available.  It uses a custom data transpose algorithm
which overlaps inter-process communication with on-processor data
rearrangement allowing the code to rely exclusively on stride 1,
on-processor FFTs (for which the code uses FFTW).

For the code's configuration we turn to a paper by
\cite{SmithLM:Gensls}. In that work they not only studied the
generation of large, slow scales in rotating and stratified flow but
they also found that for strongly stratified flows they recovered the
sheet-like structures described when Vertically Sheared Horizontal
(VSH) dynamics dominates which was discussed \cite{EmbidPF:LowFnl,
  RileyJJ:Flumps, RileyJJ:Dyntsi,BabinA:Onara}. In the spirit of those
simulations we examine some aspects of the low Rossby number limit by
using a similar simulation set-up. The principal difference between
the code set-up of \cite{SmithLM:Gensls} and this work is that instead
of using high wave number white noise forcing we use low wave number
white noise forcing which can be understood by considering the Rossby
deformation radius, $L_d$,
\begin{equation}
L_d = \frac{N}{f} L_f \qquad \text{or} \qquad k_d = \frac{f}{N} k_f.
\end{equation}
where $k_f$ is the peak wave number of the forcing and $k_d$ is the
wave number of the deformation radius. This equation shows that the
important horizontal length scales described by $k_d$ increase with
increasing rotation rate assuming $N$ is held fixed (see Table
(\ref{table:Ro256})).  Because of limited resolution we choose $k_f=3$
for all the simulations presented in this paper.

For the sake of completeness we outline some of the details of the
\cite{SmithLM:Gensls} code configuration. First, in order to reduce
the effects of viscosity in the intermediate range of scales, a
hyperviscosity replaces the Laplacian dissipation used in Eq.
(\ref{eq:bousnddim}).  The momentum dissipation is replaced by
$(-1)^{p+1} \nu_h (\nabla^2)^p \mathbf{v},$ and the buoyancy by
$(-1)^{p+1} \kappa_h (\nabla^2)^p \rho$, where $p=8$ for all the
simulations presented in this paper.  The hyperviscosity, $\nu_h$, is,
\begin{equation}
\nu_h = 2.5 \biggl(\frac{E(k_m,t)}{k_m}\biggr)^{\frac{1}{2}} ~k_m^{2-2p},
\end{equation}
as in \cite{Chasnov1994}, where $k_m$ is the highest
available wave number and $E(k_m,t)$ is the kinetic energy in the
highest available wave number shell.  The hyperdiffusivity used for
the buoyancy equation is similar.  The random forcing spectrum
$F(k)$ is Gaussian with a standard deviation $s=1$ and energy input
rate $\epsilon_f = 1$ given by,
\begin{equation}
\label{eq:forcing}
F(k) = \epsilon_f \frac{\text{exp}(-.5(k-k_f)^2/s^2)}{(2\pi)^{\frac{1}{2}}s}.
\end{equation}
In like manner we also use Rossby and Froude numbers based on the
energy input rate $\epsilon_f$ and the peak wave number $k_f$ of the
forcing,
\begin{eqnarray}
Fr = \frac{(\epsilon_f (2\pi k_f)^2)^{1/3}}{N} \qquad \text{and} \qquad Ro = \frac{(\epsilon_f
  (2\pi k_f)^2)^{1/3}}{f}.
\end{eqnarray}
The characteristic scales for time, energy, and potential enstrophy are,
\begin{eqnarray}
\label{eq:scales}
\mathcal{T} = \left(\epsilon_f (2\pi  k_f)^2\right)^{-1/3}, \qquad
\mathcal{E} = \left(\epsilon_f (2\pi k_f)^{-1}\right)^{2/3},\qquad
\text{and} \qquad
\mathcal{Q} = \rho_o^2 \left( \epsilon_f (2\pi k_f)^5 \right)^{2/3}
\end{eqnarray}
These scales are used throughout the numerical section to
nondimensionalize time, energy, and potential enstrophy. The factors
of $2\pi$ appear in the above expressions because the code has a
domain of $[0,1]^3$.

\begin{figure}
\begin{center}
\epsfig{file=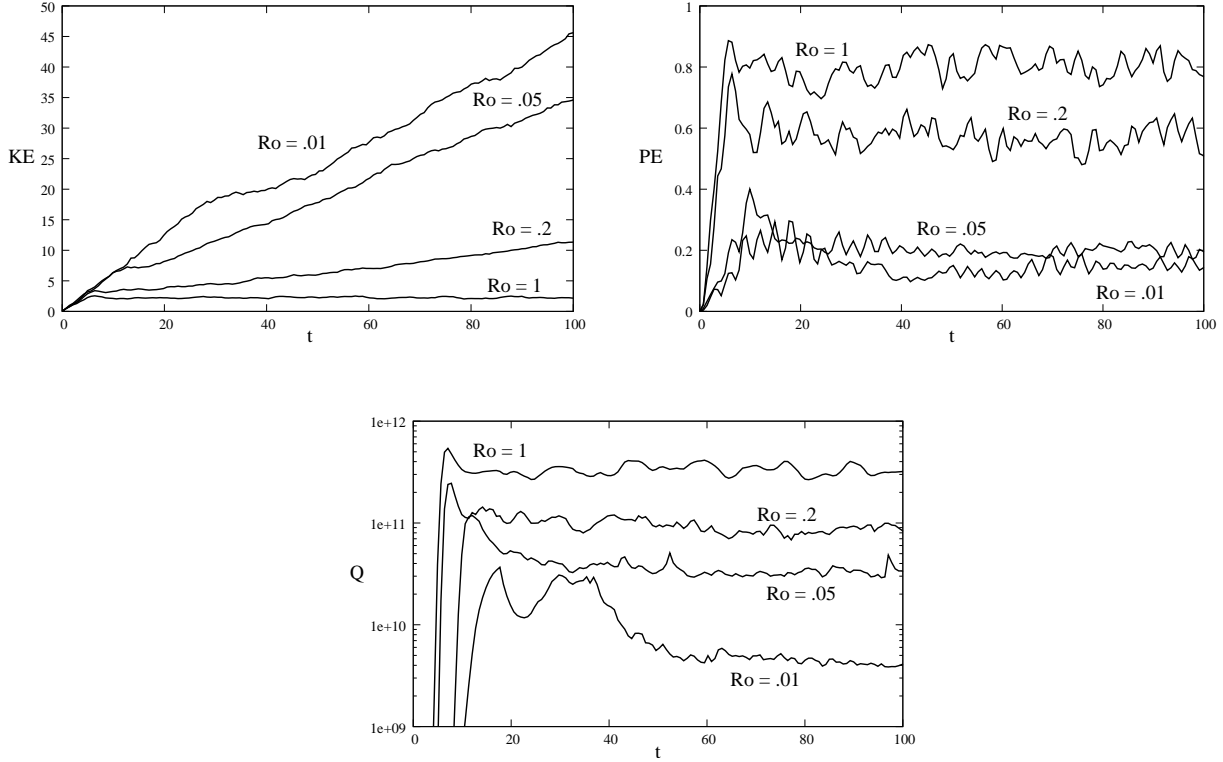}
\caption{To characterize the numerical simulations, this figure shows
  the time evolution of the kinetic energy, potential energy, and
  potential enstrophy for selected Rossby numbers and $Fr = 1$.
  Energy, potential enstrophy, and time have been nondimensionalized
  using the scales in Eq.  (\ref{eq:scales}). The kinetic energy grows
  in time due to the momentum forcing. After a spin-up time,
  comparison of the different runs shows that the smaller the Rossby
  number the larger the magnitude of the kinetic energy, and the
  smaller the magnitude of the potential energy. The order of
  magnitude of the potential enstrophy also decreases with decreasing
  Rossby number. For all the simulations except $Ro=1$ columns appear
  that span the depth of the fluid. These columns are dynamic and
  remain columnar for the duration of the simulation.
}
\label{fig:totalenvst}
\end{center}
\end{figure}

Each run is spun up from zero and forced throughout by the low wave
number white noise forcing described by Eq. (\ref{eq:forcing}). The
parameters $N$ and $f$ remain fixed throughout the simulation. For
runs where ($Ro < 1$) columnar structures form during the spin up
period.  The columns are dynamic and retain their basic columnar
structure in the quantities $\overline{u}^z$ and $\overline{u}^x$
throughout the simulation (see Fig. (\ref{fig:averages2})).

The evolution of the kinetic energy, potential energy, and potential
enstrophy, nondimensionalized using the scales in Eq.
(\ref{eq:scales}), for selected Rossby numbers are show in Fig.
(\ref{fig:totalenvst}).  Because the flow is forced in the momentum
equations the kinetic energy increases with time.  After a spin-up time,
comparison of the different runs shows that the smaller the Rossby
number the larger the magnitude of the kinetic energy, and the smaller
the magnitude of the potential energy. The order of magnitude of the
potential enstrophy also decreases with decreasing Rossby number.

\subsection{Columnar Taylor-Proudman flows} The classical
Taylor-Proudman result is that for constant density flow in
geostrophic and hydrostatic balance the vertical derivatives of the
horizontal and vertical velocities are zero, creating columnar flows.
Our theory, described by the projection operator, Eq.
(\ref{eq:projontonull}), generalizes the Taylor-Proudman theory to the
case when the density is not constant and the flow is not in
hydrostatic balance. We examine the flow characteristics at different
Rossby numbers to see if Taylor-Proudman flows appear.  First we
define two averages of $u$, the $x$ component of the horizontal velocity,
\begin{figure}
\begin{center}
\epsfig{file=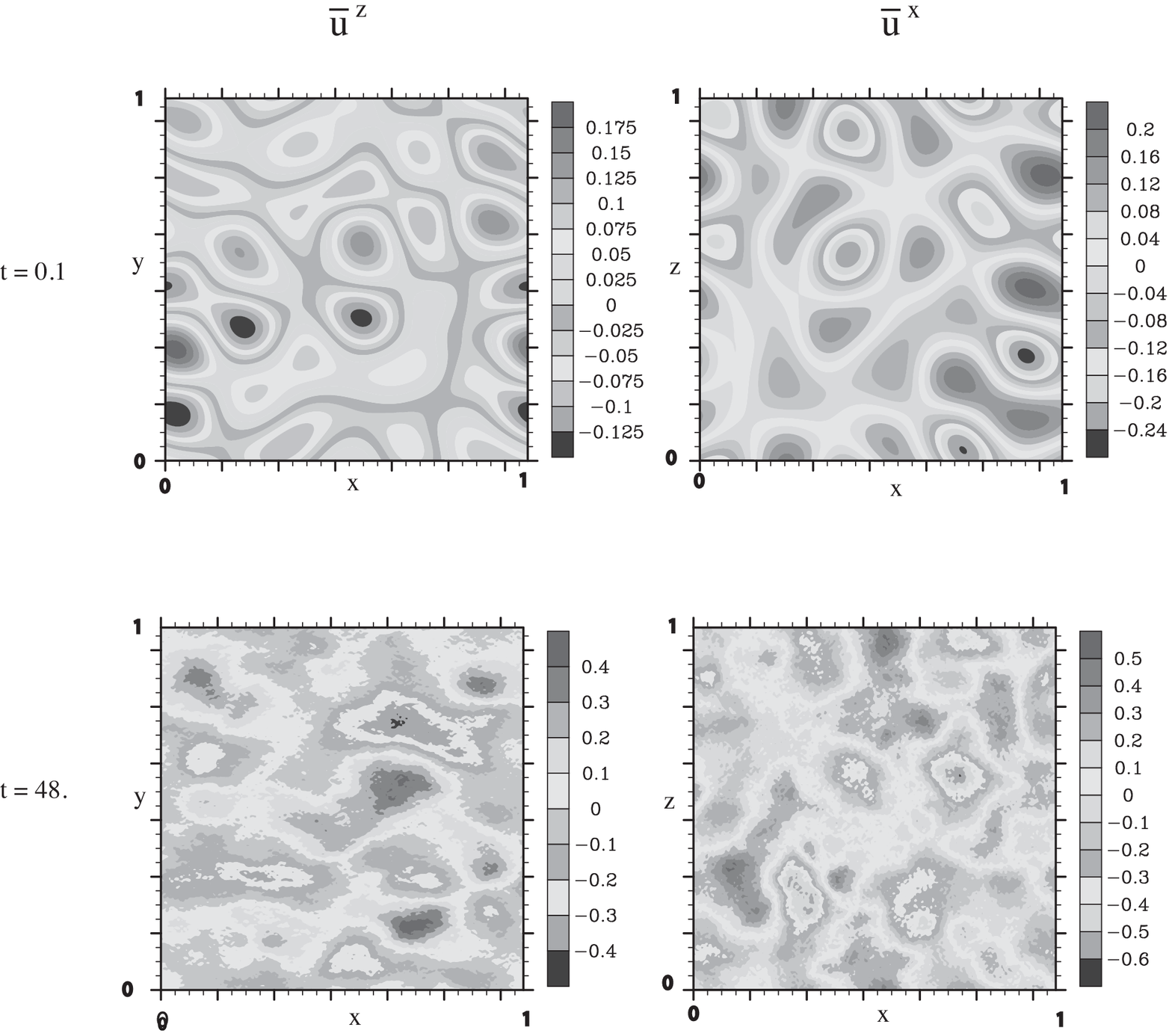,height=7.0in}
\caption{This figure shows vertical (left side) and horizontal (right
  side) averages of the horizontal component of the velocities for $Ro
  = 1$ and $Fr = 1$ simulations of the full Boussinesq equations.
  Patterns form on length scales consistent with the low wave number
  forcing of $k_f=3$ but no columns form.}
\label{fig:averages1}
\end{center}
\end{figure}

\begin{figure}
\begin{center}
\epsfig{file=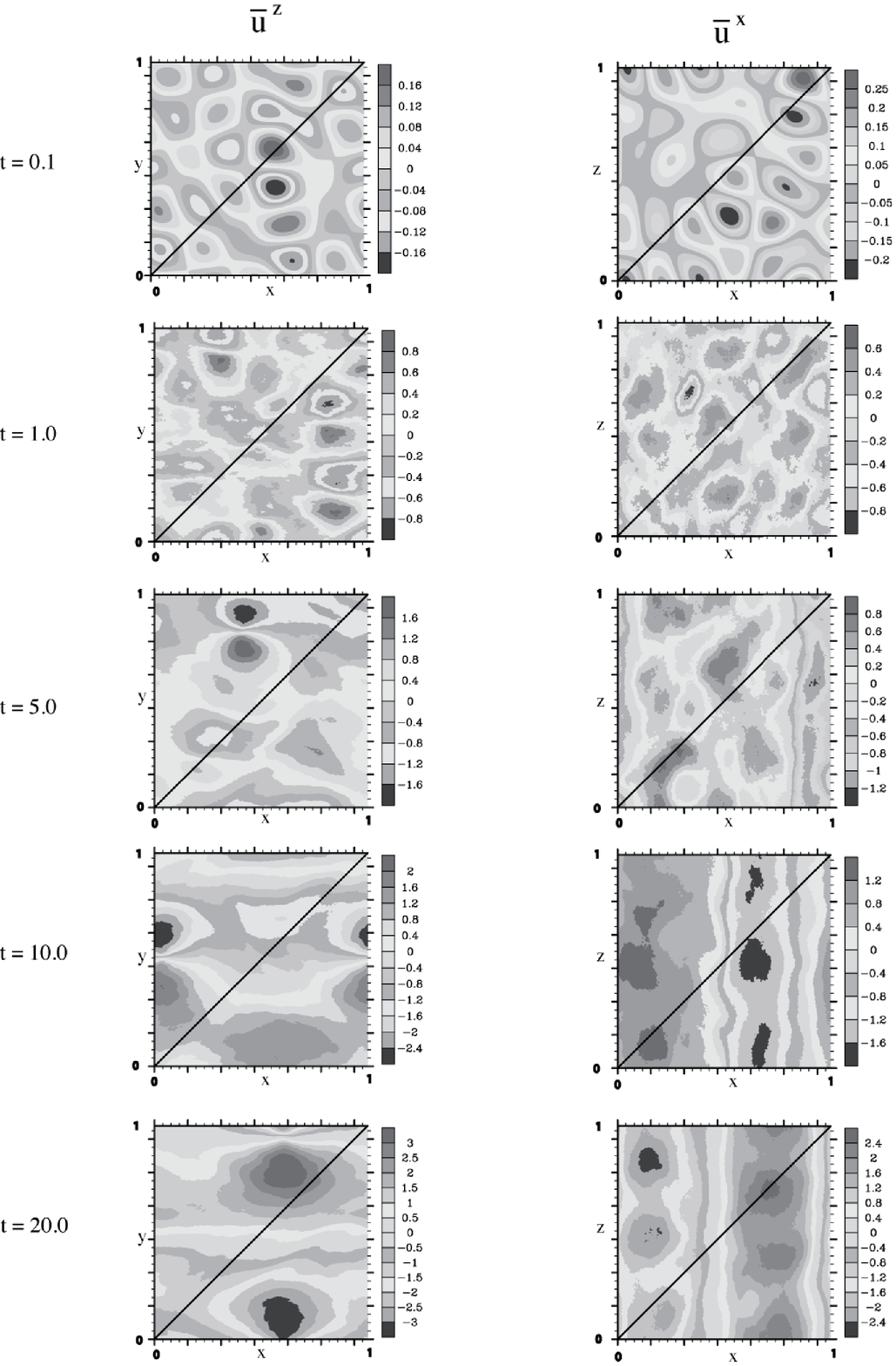,height=7in}
\caption{This figure shows vertical (left column) and horizontal
  (right column) averages of the horizontal component of the
  velocities for $Ro = .2$ and $Fr = 1$ simulations of the full
  Boussinesq equations.  By time $t=10$ columnar structures are
  beginning to form. At $t=20$ the dominant columnar structures have
  formed and remain columnar through the rest of the simulation.}
\label{fig:averages2}
\end{center}
\end{figure}

\begin{eqnarray}
\overline{u}^z = \int_L u(x,y,z,t_o) 2\pi~dz, \qquad \text{and} \qquad
\overline{u}^x = \int_L u(x,y,z,t_o) 2\pi~dx,
\end{eqnarray}
where $L=1$ and $t_o$ is any time after the spin up time. Fig.
(\ref{fig:averages1}) and Fig. (\ref{fig:averages2}) show contour
plots of $\overline{u}^z$, which we used to identify large-scale
horizontal structures, and contour plots of $\overline{u}^x$, which we
use to identify large-scale vertical structures.  The left panel of
each figure shows contours of $\overline{u}^z$ while the right panel
shows contours of $\overline{u}^x$. Fig. (\ref{fig:averages1}) shows
the case when $Ro=1$ (not small).  The main structures of this flow
are consistent with $k_f=3$ forcing scale but never become
columnar flows.  On the right of Fig. (\ref{fig:averages2}), for the
case of a smaller Rossby number, $Ro = .2$, the contours reveal a low
wave number vortical structure.  That this is a columnar structure
becomes evident by examining the lower right panel which shows the
contour plot of $\overline{u}^x$.  There is no similar columnar
structure for the $Ro=1$ run.  Therefore, even with low wave number
white noise forcing columnar Taylor-Proudman flows {\sl spontaneously
  form} if the Rossby number is low enough.

Though much is known about the formation and instability of vortices
in rotating and stratified flow, this is the first study that we know
of where a constant-in-time white noise forcing creates columnar
structures that remain columnar throughout the length of the
simulation.

Other studies that discuss the formation of columnar structures in
rotating turbulence (with no buoyancy) can be found in
\cite{DavidsonStaplehurstDalziel2006, StaplehurstDavidsonDalziel2008,
  SreenivasanDavidson2008}. These are studies about the creation and
evolution of vortices from an initial condition, unlike this work
which has a constant-in-time white noise forcing, and they do not
consider stratification, while this work considers only weak
stratification. Despite this, the mechanisms they found for columnar
vortex formation are relevant to this study.  In
\cite{DavidsonStaplehurstDalziel2006, StaplehurstDavidsonDalziel2008}
they begin by considering an initial blob of fluid in a rotating
environment.  They find that when the rotation is strong enough linear
wave energy propagation is biased along the axes of rotation and that
when the columnar vortex appears it remains contained in the cylinder
that circumscribes the initial blob. When the rotation is weak a
centrifugal bursting phenomenon prevents any columnar vortices from
forming. We can see evidence for this in Fig.  (\ref{fig:averages2})
as the wave number 3 structures elongate and finally form columnar
structures. There are two classical laboratory studies of the
formation of columnar vortices and both employ a constant-in-time
forcing: the original laboratory studies by \cite{Taylor} and
\cite{Davies72}.  Both investigators studied the dynamics associated
with the slow, steady, horizontal motion of a solid obstacle through a
fluid rotating about a vertical axis. One main difference between the
studies is that \cite{Davies72} included stratification and
\cite{Taylor} did not. Both investigators find the formation of
columnar structures above the moving topographic feature. While there
are considerable differences in the kinds of columnar vortices that
form, when they did form there was no mention of them becoming
unstable.

There is also a large body of literature on the stability of columnar
vortices in rotating and stratified flow. We restrict this discussion
to key work related to this study (strong rotation and weak
stratification, i.e. nonhydrostatic). We first consider the work of
\cite{PotylitsinPeltier1998} in which they use linear stability
analysis to study the stability of columnar vortices to
three-dimensional perturbations. They use two initial distributions of
vorticity: Kelvin-Helmholtz-generated vortices in shear and Kida-like
vortices in strain. Their conclusion is that an isolated anticyclonic
vortex column is strongly destabilized by small values of the
background rotation, while rapid rotation stabilizes both cyclonic and
anticyclonic initial conditions. They explain this phenomenon using
the Taylor-Proudman theorem. They also discuss the details of the
stability of anticylonic vortices but since all our columnar vortices
are cyclonic we will not describe their other results here. The next
related paper is \cite{PotylitsinPeltier2003} where they use direct
numerical simulations to study the evolution of
Kelvin-Helmholtz-generated columnar vortices and verify their previous
results that strong rotation stabilizes the columnar vortices. The
last work we will examine is \cite{Otheguy2006}. This work also uses a
linear stability analysis but begins with an initial condition of
co-rotating vortices. When there is no stratification they find, as
\cite{PotylitsinPeltier1998}, that the columnar vortex is elliptic
unstable and that stronger stratification causes a zig-zag
instability. This work is not directly applicable to our work because
it studies the evolution of an initial condition of co-rotating
vortices which our simulations do not show but it shows that the
evolution of columnar vortices depends on the rotation,
stratification, initial conditions and forcing.

\subsection{Ratio of slow energy to total energy} 
\begin{figure}
\begin{center}
\epsfig{file=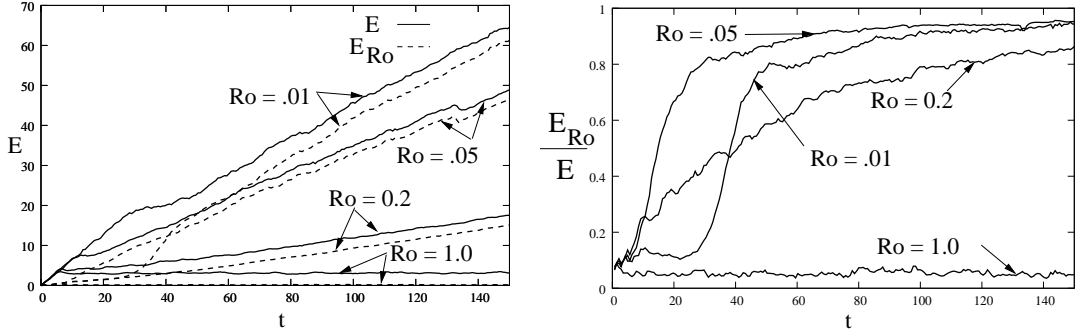}
\caption{The panel on the left shows the total energy, $E$ (solid
  line) and the slow total energy, $E_{Ro}$ (dashed line). The total
  energy and the slow total energy appear to be on parallel
  trajectories in time.  This is explored more fully in the panel on
  the right which shows the time evolution of the ratio $E_{Ro}/E$. As
  the Rossby number decreases a larger fraction of the total energy is
  slow.  Furthermore, the low Rossby number runs appear to be
  approaching a constant at late times.}
\label{fig:totewithtime}
\end{center}
\end{figure}

Our theory states that in the absence of viscosity the total energy is
composed of both fast and slow dynamics but that the ratio of the $Ro
\rightarrow 0$ (slow) total energy, $E_{Ro}$, to the total energy, $E$,
should go to a constant ( cf. the discussion following Eq. (\ref{eq:fastenergy}))
\begin{equation}
    \frac{E_{Ro}}{E} \rightarrow C \qquad \text{for} \qquad ~\nu = 0.
  \end{equation}
  In this section we examine the time evolution of $E_{Ro}/E$ of our
  numerical simulations for varying Rossby numbers. The total energy,
  $E$, is given by,
\begin{equation}
  \label{eq:energyndt}
  E =  \int_V~ \frac{1}{2}\left(|\mathbf{v}|^2 + \rho^2 \right) (2\pi)^3 ~d
  \slv,
\end{equation} 
where $V = 1$. The $Ro \rightarrow 0$ (slow) component is
computed by projecting the full solution vector, $(u,v,w,\rho)$ onto
the null space of the fast operator using Eq. (\ref{eq:projontonull}).
The total energy from the slow variables is then computed by combining
the slow horizontal kinetic energy described by Eq. (\ref{eq:2dhke})
and the total vertical energy described in Eq. (\ref{eq:2dve}),
\begin{equation}
  \label{eq:energyndtt}
  E_{Ro} = ~\int_A \frac{1}{2} \left( ~|\mathbf{v}_H|^2  +  w^2 +
  \avez{\rho}^2 \right) (2\pi)^2 d \sla,
\end{equation}
where $A = 1$ and, as mentioned above, all the variables are
the result of projecting the full solution vector onto the null space
of the fast operator.  The left panel of Figure
(\ref{fig:totewithtime}) shows the evolution of both $E$ (solid line)
and $E_{Ro}$ (dashed line) with time.  At fixed time, both the slow
total energy and the total energy have larger amplitudes as the Rossby
number decreases.  The panel also shows that the total energy and the
slow total energy appear to maintain a similar ratio as time
increases.  This is explored more fully in the right panel where we
plot the ratio $E_{Ro}/E$. We find that when $Ro = 1$ the ratio
$E_{Ro}/E$ maintains approximately the same percentage of slow to
total energy, but that the slow component is a small fraction of the
total. As we decrease the Rossby number the time evolution of
$E_{Ro}/E$ shows a gradual increase in the percentage of energy that
is slow relative to the total.  For the smallest Rossby numbers we
ran, the ratio quickly increases to a value where a substantial
fraction of the total energy is its slow component and then gradually
increases toward a constant close to one. While $E_{Ro}/E$ is not a
constant in time, it is nearly so for the smallest Rossby numbers.
This suggests that even when there is dissipation and forcing in the
system there is a rapid initial adjustment of $E_{Ro}/E$ toward a
value that indicates a significant fraction of the total energy is
slow, then a gradual increase toward a constant close to one.

\begin{figure}
\begin{center}
\epsfig{file=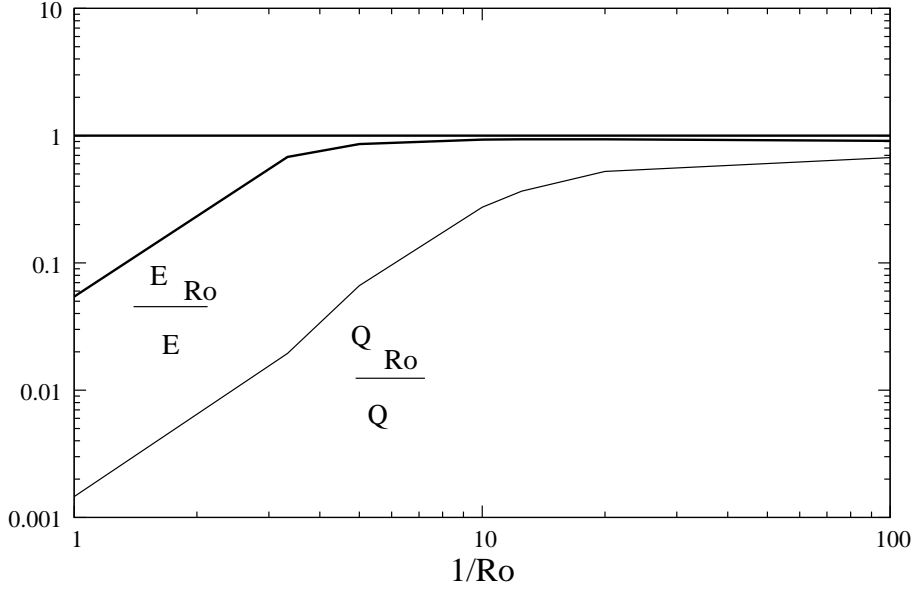}
\caption{This figure shows the dependence of the time averaged
  quantities $\overline{E}_{Ro}^T/\overline{E}^T$ and
  $\overline{Q}_{Ro}^T/\overline{Q}^T$ on $Ro$. As the Rossby number
  decreases the simulations show that a larger fraction of the total
  energy and potential enstrophy is slow. }
\label{fig:qratiovsro}
\end{center}
\end{figure}

To examine the explicit dependence of $E_{Ro}/E$ on Rossby
number we compute its time average, 
$\overline{E}_{Ro}^T/\overline{E}^T$, where we compute the average using,
\begin{equation}
\label{eq:timeavg}
\overline{X}^T = \frac{1}{T_F-T_0}\sum_{i = T_0}^{T_F} X_i ~dt_i
\end{equation}
where $T_0$ is a time immediately after spin-up and $T_F$ is the last
time available from the simulation. This quantity is shown in Fig.
(\ref{fig:qratiovsro}) where we see the trend that as the Rossby
number decreases, the ratio of slow total energy to total energy is
approaching a constant and that the constant is close to one. This
implies that at the low Rossby numbers most of the total energy in
these simulations is slow.

\subsection{Ratio of slow potential enstrophy to total potential
  enstrophy} Our theory shows that in the limit of $Ro \rightarrow 0$ and 
$Fr = O(1)$,  the potential enstrophy contains only slow dynamics (see the discussion
below Eq. (3.41).  Stated another way, the ratio
of the slow potential enstrophy, $Q_{Ro}$, to the total potential
enstrophy, $Q$, goes to one,
  \begin{equation}
    \frac{Q_{Ro}}{Q} \rightarrow 1 \qquad \text{for} \qquad~\nu = 0.
  \end{equation}
  Here we examine the time evolution of $Q_{Ro}/Q$ for our
  numerical simulations for a sequence of decreasing Rossby numbers.
  We compute the total potential enstrophy as,
\begin{equation}
Q = \int_V ~\frac{1}{2} q^2 ~d \slv
\label{eq:totpotensvol}
\end{equation}
where $V = (2\pi)^3$, and $q$ is the nondimensional potential
vorticity described by Eq. (\ref{eq:potvorndexpand}). The $Ro
\rightarrow 0$ component of potential enstrophy is computed by
projecting the full solution vector onto the null space of the fast
operator, Eq. (\ref{eq:projontonull}). However, since the projection
operator does not affect $\rho$, the $Ro \rightarrow 0$ component
of the potential enstrophy can be computed by,
\begin{equation}
Q_{Ro} = \int_V \frac{1}{2} \left( \frac{\partial\rho}{\partial
    z}\right)^2 ~d \slv.
\label{eq:totpotensvolslow}
\end{equation}

\begin{figure}
\begin{center}
\epsfig{file=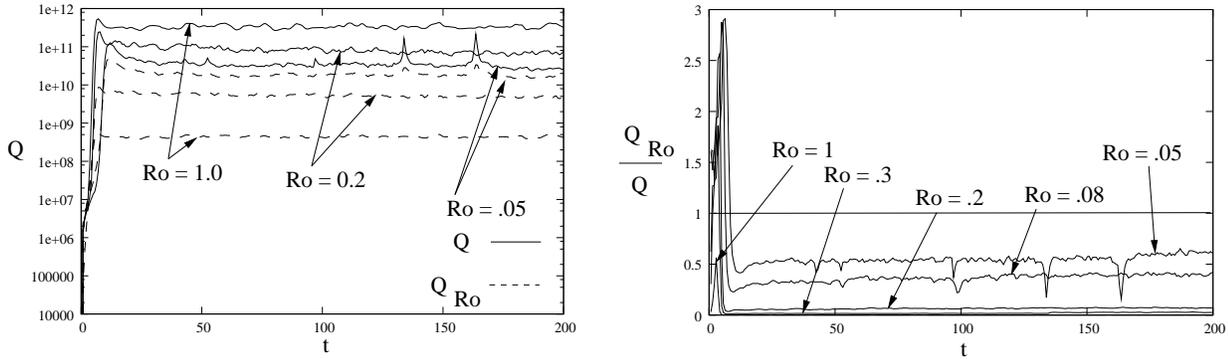}
\caption{The left panel shows the time evolution of the potential
  enstrophy (solid line) and the slow potential enstrophy (dashed
  line). As the Rossby number decreases the curves that represent the
  slow potential enstrophy and the total potential enstrophy converge
  toward each other.  This is explored more fully in the right panel
  which shows the time evolution of the ratio $Q_{Ro}/Q$. For the
  lowest Rossby number, $Ro = .01$, the magnitude of the ratio starts
  to decrease after about $T=30$ indicating that the slow component of
  the potential enstrophy decreases. However, at that Rossby number
  the simulations are approaching the limit of what $256^3$
  simulations can resolve.  For the other Rossby numbers, as $Ro$
  decreases this ratio approaches one, showing that in the limit of
  fast rotation and order one stratification a large fraction of their
  potential enstrophy is its slow component - even with white noise
  forcing.}
\label{fig:potensvst}
\end{center}
\end{figure}

The panel on the left of Figure (\ref{fig:potensvst}) compares the
time evolution of $Q$ (solid lines) with the time evolution of
$Q_{Ro}$ (dashed lines). As the Rossby number decreases, the gap
between the solid and dashed lines decreases, indicating that the
total enstrophy's composition has a larger slow component.  This is
explored further in the right panel where we show the ratio
$Q_{Ro}/Q$. As the Rossby number tends to smaller values the slow
component of the potential enstrophy becomes a larger fraction of the
total potential enstrophy. This suggests that even with low wave
number white noise forcing the potential enstrophy of this limit is
dominated by its slow component. We also note that as the Rossby
number decreases the vorticity is expected to have a larger fast
component (see Section \ref{subsec:fastlimit}) which means the
potential enstrophy defined by Eq. (\ref{eq:totpotensvol}) can be
replaced with Eq. (\ref{eq:totpotensvolslow}).

Finally, to see the dependence of the ratio $Q_{Ro}/Q$ on the Rossby
number, we plot the quantity $\overline{Q}_{Ro}^T/\overline{Q}^T$
where the averages are computed using Eq. (\ref{eq:timeavg}) and are
shown in Fig.  (\ref{fig:qratiovsro}). This figure shows that
$Q_{Ro}/Q \rightarrow 1$ as $Ro$ decreases the ratio is approaching
one but more slowly in Rossby number than that of the energy.
Dimensionally this means the globally integrated potential enstrophy
is dominated by the vertical gradient of the buoyancy times the
Coriolis parameter while both the vorticity times the Brunt-V\"ais\"ala frequency
and the nonlinear terms have lesser influence.

\section{Summary}

We have examined the fast rotation limit of the rotating and
stratified Boussinesq equations using the framework of
\cite{EmbidPF:LowFnl}. We have shown that to leading order, the
dynamics is composed of both fast and slow components but that the
slow dynamics evolves independently of the fast. We have also derived
new equations for the slow dynamics. These include the two dimensional
Navier-Stokes equations for the slow horizontal velocity, a forced
advection-diffusion equation for the vertically averaged vertical
velocity, making the slow dynamics {\sl{non-hydrostatic}}, and an
equation for the buoyancy which is the only quantity to retain its
three-dimensional character. In the absence of viscosity and
diffusivity these new sets of equations conserve a horizontal kinetic
energy and vertical vorticity, along with a new conserved quantity
that describes dynamics between the vertical kinetic energy and the
buoyancy. The leading order total energy contains both fast and slow
dynamics, though their ratio is conserved. The potential energy is
found, to leading order, to contain only slow dynamics.

Our numerical simulations, which used low wave number white noise
forcing in the momentum equations, reveal the emergence of
Taylor-Proudman flows as the Rossby number decreases. They also
support the theory that the ratio of the slow to total energy goes to
a constant as the Rossby number is decreased and that the constant is
close to one. In addition to the energy we also examined the ratio of
the slow potential enstrophy to the total potential enstrophy. The
smallest Rossby number we examined, $Ro = .01$, was at the limit of
the scales that could be resolved at this resolution. However, the
other Rossby number numbers showed the trend that the ratio of the
slow potential enstrophy to total potential enstrophy also approaches
a constant and that constant trends toward one.  These numerical
simulations indicate that some of the aspects dynamics derived in this
paper exist even in the presence of white noise forcing and
hyperviscosity.

\setcounter{equation}{0}
\renewcommand{\theequation}{\mbox{A.} \arabic{equation}}
\subsection*{Appendix}
\subsection*{Analysis of the fast operator}
The fast operator $L_F$ is defined on the Hilbert space $X$ of $2\pi$-periodic
square-integrable vector fields $\bfu = (\bfv , \rho)$ that are divergence
free, $\grad \cdot \bfv = 0$. In the space $X$ the eigenfunctions of $L_F$ are
given by Fourier modes of the form 
$\bfu_\bfk (\bfx) = e^{i \bfk \cdot \bfx} \bfr_\bfk$ where $\bfk = (k,l,m)$ is 
the wave number and $\bfr_\bfk = (\bfv_\bfk, \rho_\bfk)$ is a fixed 
vector. The divergence free condition reduces to the algebraic constraint
$\bfv_\bfk \cdot \bfk = 0$. In terms of the Fourier eigenmode 
$\bfu_\bfk ( \bfx)$ the eigenvalue equation 
$L_F \bfu_\bfk = \lambda_\bfk \bfu_\bfk$ reduces to the algebraic
eigenvalue problem $L_F(\bfk) \bfr_\bfk = \lambda_\bfk \bfr_\bfk$, where 
the matrix symbol $L_F(\bfk)$ is given, for $\bfk \neq \bfO$ and 
$\bfk = \bfO$ respectively, by 
\begin{equation}
L_F(\mathbf{k}) = \frac{1}{|\bfk|^2}
\begin{pmatrix}
-k l & -(l^2+m^2) & 0 & 0 \\
k^2+m^2 & k l & 0 & 0 \\
-l m & k m & 0 & 0 \\
0 & 0 & 0 & 0
\end{pmatrix},  \ \
L_F(\bfO) = 
\begin{pmatrix}
0 & -1 & 0 & 0 \\
1 & 0 & 0 & 0 \\
0 & 0  & 0 & 0 \\
0 & 0 & 0 & 0
\end{pmatrix}, 
\end{equation}
with $|\bfk|^2 = k^2 + l^2 + m^2$. The algebraic eigenvalue problem has 
four purely imaginary eigenvalues, $\lambda_\bfk = i \w_\bfk^\za$, with 
$\w_\bfk^{\pm 1} = \pm \frac{m}{|\bfk|}$ corresponding to  fast 
inertial modes and the double eigenvalue $\w_\bfk^0 = 0$ corresponding 
to the slow modes. The associated eigenvectors are given as follows. If
$|\bfk_H| \neq 0$ there are three eigenvectors,
\begin{equation}
\label{eq:eigv1}
\bfr^1_\bfk = \frac{1}{\sqrt{2} | \bfk_H| | \bfk |} 
\begin{pmatrix}
-l |\bfk| + i k m \\
k |\bfk| + i l m  \\
-i|\bfk_H|^2  \\
0
\end{pmatrix}, \ \
\bfr^{-1}_\bfk = \frac{1}{\sqrt{2} | \bfk_H| | \bfk |} 
\begin{pmatrix}
l |\bfk| - i k m \\
-k |\bfk| - i l m \\
i|\bfk_H|^2 \\
0
\end{pmatrix}, \ \
\bfr^0_\bfk = 
\begin{pmatrix}
0 \\
0 \\
0 \\
1
\end{pmatrix} .
\end{equation}
The fourth eigenvector does not satisfy the incompressibility constraint
$\bfv_\bfk \cdot \bfk = 0$. If  $|\bfk_H| = 0$ but $|\bfk| \neq 0$, 
then $\w_\bfk^{\pm 1} = \pm 1$, $\w_\bfk^0 = 0$, and there are 
four eigenvectors,
\begin{equation}
\label{eq:eigv2}
\bfr^1_\bfk = \frac{1}{\sqrt{2}} 
\begin{pmatrix}
1 \\
-i \\
0 \\
0 
\end{pmatrix}, \ \
\bfr^{-1}_\bfk = \frac{1}{\sqrt{2}} 
\begin{pmatrix}
1 \\
i \\
0 \\
0 
\end{pmatrix}, \ \
\bfr^0_\bfk = 
\begin{pmatrix}
0 \\
0 \\
0 \\
1
\end{pmatrix}, \ \
\tilde{\bfr}^0_\bfk =
\begin{pmatrix}
0 \\
0 \\
1 \\
0
\end{pmatrix}, 
\end{equation}
but the fourth eigenvector, $\tilde{\bfr}^0_\bfk$, violates the incompressibility 
constraint. Finally, if $|\bfk| = 0$ then there are two fast modes associated
with $\w_\bfO^{\pm 1} = \pm 1$ and two slow modes associated with 
$\w_\bfO^0 = 0$, with the four eigenvectors in Eq. (\ref{eq:eigv2}) now 
satisfying the incompressibility constraint. 
Notice that the eigenfunctions are normalized and satisfy the symmetry 
condition $\overline{\bfr_\bfk^\za} = \bfr_{-\bfk}^{-\za}$, where the bar stands
for complex conjugation. For this reason we require that the amplitudes 
$\sigma_\bfk^\za (t)$ in Eq. (\ref{eq:eigenexpansion}) satisfy the condition 
$\overline{\sigma_\bfk^\za} = \sigma_{-\bfk}^{-\za}$ to ensure that $\bfu(\bfx, t)$ 
in Eq. (\ref{eq:eigenexpansion}) is real valued. 
\subsection*{Formulas for the interaction coefficients}
Here we collect the formulas for the interaction coefficients 
$B^{(\za',\za'',\za)}_{(\bfk',\bfk'',\bfk)}$, $L_\bfk^{(\za',\za)}$ and 
$D_\bfk^{(\za',\za)}$, which appear in Fourier formulation of the limiting
fast dynamics equations, Eq. (\ref{eq:fastcoef}).  
The quadratic interaction coefficient $B^{(\za',\za'',\za)}_{(\bfk',\bfk'',\bfk)}$
is given by,
\begin{eqnarray}
\label{eq:Bcoef}
&&B^{(\alpha',\alpha'',\alpha)}_{(\bfk',\bfk'',\bfk)} = \frac{i}{2} \left[
(\bfv_{\bfk'}^{\za'} \cdot \bfk'') \bfr_{\bfk''}^{\za''} +
(\bfv_{\bfk''}^{\za''}\cdot \bfk') \bfr_{\bfk'}^{\za'} 
\right] \cdot \overline{\mathbf{r}_{\bfk}^{\alpha}}. 
\end{eqnarray}
With this formula we can verify the claim that the fast limiting dynamics 
equations for the slow modes is independent of the fast modes, i.e. that 
the quadratic interaction coefficients corresponding to 
``fast + fast $\rightarrow$ slow'' are zero. Because the formula
for the quadratic interaction coefficient in Eq. (\ref{eq:Bcoef}) is 
invariant under the  permutation of $\za'$ and $\za''$, and $\bfk'$ and $\bfk''$,
it is sufficient to check that $B^{(-1,1,0)}_{(\bfk',\bfk'',\bfk)}$ is zero.
The three-wave resonance equations for this case is
\begin{equation}
\label{eq:resonance}
\bfk' + \bfk'' = \bfk, \quad \frac{m'}{|\bfk'|} - \frac{m''}{|\bfk''|} = 0.
\end{equation}
There are three cases to consider. First, if $|\bfk_H| \neq 0$  then 
$\bfr_\bfk^0$ in Eq. (\ref{eq:eigv1}) is orthogonal to $\bfr_\bfk^{\pm 1}$
in both Eq. (\ref{eq:eigv1}) and Eq. (\ref{eq:eigv2}), and 
$B^{(-1,1,0)}_{(\bfk',\bfk'',\bfk)}$ is zero . Second, if $|\bfk_H| = 0$ but
$|\bfk| \neq 0$ then $\bfr_\bfk^0$ in Eq. (\ref{eq:eigv2}) coincides
with $\bfr_\bfk^0$ in the previous case and the quadratic coefficient is 
again zero. Finally, if $|\bfk|=0$ then there $\bfr_\bfk^\za$ in
Eq. (\ref{eq:eigv2}) is either $\bfr_\bfk^0$ or $\tilde{\bfr}_\bfk^0$.  
In this case $\bfk' = -\bfk''$ and, by symmetry, 
$\bfr_{-\bfk'}^{-1} = \overline{\bfr_{\bfk'}^1}$. Direct calculation then shows
that the third component of 
$(\bfv_{\bfk'}^1 \cdot (-\bfk')) \overline{\bfr_{\bfk'}^1} +
(\overline{\bfv_{\bfk'}^1}\cdot \bfk') \bfr_{\bfk'}^1$ in 
Eq. (\ref{eq:Bcoef}) is given by,
\begin{eqnarray}
\nonumber
(\bfv_{\bfk'}^1 \cdot (-\bfk')) \overline{w_{\bfk'}^1} +
(\overline{\bfv_{\bfk'}^1}\cdot \bfk') w_{\bfk'}^1  &=&
i |\bfk_H'|^2( \bfv_{\bfk'}^1 + \overline{ \bfv_{\bfk'}^1}) \cdot \bfk' \\ 
&=&
2i |\bfk_H'|^2 (-l' |\bfk'|, k'|\bfk'|, 0) \cdot (k',l',m') = 0, 
\end{eqnarray}
and this implies that the dot product with either $\bfr_\bfk^0$ or 
$\tilde{\bfr}_\bfk^0$ for the quadratic interaction coefficient in
Eq. (\ref{eq:Bcoef}) is again zero.  This proves that 
$B^{(-1,1,0)}_{(\bfk',\bfk'',\bfk)}$ is always zero.

Next, the linear interaction coefficient $L_\bfk^{(\za',\za)}$
in Eq. (\ref{eq:fastcoef}) is given by,
\begin{equation}
L_\bfk^{(\alpha',\alpha)} = (\bfr^\za_\bfk)^*  L_S(\mathbf{k})
\mathbf{r}_{\bfk}^{\alpha'}, 
\end{equation}
where the matrix symbol  $L_S(\bfk)$ associated with the slow operator $L_S$
is given, for $\bfk \neq \bfO$ and $\bfk = \bfO$ respectively, by  
\begin{equation}
L_S(\mathbf{k}) = \frac{1}{Fr}
\begin{pmatrix}
0 & 0 & 0 & - \frac{k m}{|\bfk|^2}\\
0 & 0 & 0 & - \frac{l m}{|\bfk|^2}\\
0 & 0 & 0 & \frac{|\bfk_H|^2}{|\bfk|^2}\\
0 & 0 & -1 & 0
\end{pmatrix},  \quad   
L_S(\bfO) = \frac{1}{Fr}
\begin{pmatrix}
0 & 0 & 0 & 0 \\
0 & 0 & 0 & 0 \\
0 & 0 & 0 & 1 \\
0 & 0 & -1 & 0
\end{pmatrix}.
\end{equation}
Direct calculation of these coefficients with the eigenvectors 
given in Eqs. (\ref{eq:eigv1}) -- (\ref{eq:eigv2}) gives the 
explicit values of $L_\bfk^{(0, \pm 1)} = \pm i/\sqrt{2}$ and
$L_\bfk^{(\pm 1, 0)} = \mp i/\sqrt{2}$ when $\bfk = (k,l,0)$,
$L_{\bf0}^{(0,\tilde{0})}= 1$, and $L_{\bf0}^{(\tilde{0},0)} = -1$ when
$\bfk = \bf0$, and zero otherwise.

Finally, the diffusion coefficient  $D_\bfk^{(\za',\za)}$
in Eq. (\ref{eq:fastcoef}) is given by,
\begin{equation}
D_\bfk^{(\alpha',\alpha)} = (\bfr_\bfk^\za)^* D(\bfk) \bfr_{\bfk}^{\za'},
\end{equation}
where $D(\bfk)$ is the diagonal matrix 
\begin{equation}
D(\bfk) = \mbox{diag} 
\begin{pmatrix}
 -\frac{1}{Re}|\bfk|^2, -\frac{1}{Re}|\bfk|^2,
-\frac{1}{Re}|\bfk|^2, -\frac{1}{Re Pr}|\bfk|^2  
\end{pmatrix},
\end{equation}
and direct calculation with the eigenvectors given in 
Eqs. (\ref{eq:eigv1}) -- (\ref{eq:eigv2}) shows that 
\begin{equation}
D_\bfk^{(1,1)} = D_\bfk^{(-1,-1)} = -\frac{1}{Re}|\bfk|^2, \quad  
D_\bfk^{(0,0)} =  -\frac{1}{Re Pr} |\bfk|^2, 
\end{equation}
and zero otherwise.


\rem{

}

\end{document}